\documentclass{aastex}          
\usepackage{spr-astr-addons}    
\usepackage[colorlinks,citecolor=blue]{hyperref}
\usepackage{amssymb}

\newcommand{\sfrd}{$\Sigma_{\rm SFR}$}
\newcommand{\usfr}{$\rm M_{\odot}\,yr^{-1}$}
\newcommand{\usfrd}{$\rm M_{\odot}\,yr^{-1}\,kpc^{-2}$}
\newcommand{\uvel}{$\rm km\,s^{-1}$}
\newcommand{\udif}{$\rm cm^2\,s^{-1}$}

\begin{document}
%
\title{The radio continuum perspective on cosmic-ray transport in external galaxies}

\shorttitle{Cosmic-ray transport}
\shortauthors{V. Heesen}

\author{Volker Heesen\altaffilmark{1}} 

\altaffiltext{1}{Hamburger Sternwarte, Universit\"at Hamburg, Gojenbergsweg 112, D-21029 Hamburg, Germany}

\begin{abstract}
Radio continuum observations of external galaxies provide us with an excellent outside view on the distribution of cosmic-ray electrons in the disc and halo. In this review, we summarise the current state of what we have learned from modelling such observations with cosmic-ray transport, paying particular attention to the question to what extent we can exploit radio haloes when studying galactic winds. We have developed the user-friendly framework {\sc spinnaker} to model radio haloes with either pure advection or diffusion, allowing us to study both diffusion coefficients and advection speeds in nearby galaxies. Using these models, we show that we can identify galaxies with winds using both morphology  and radio spectral indices of radio haloes. Advective radio haloes are ubiquitous, indicating that already fairly low values of the star formation rate (SFR) surface density ($\Sigma_{\rm SFR}$) can trigger galactic winds. The advection speeds scale with SFR, $\Sigma_{\rm SFR}$, and rotation speed as expected for stellar feedback-driven winds. Accelerating winds are in agreement with our radio spectral index data, but this is sensitive to the magnetic field parametrisation, so that constant wind speeds cannot be ruled out either. The question to what extent cosmic rays can be a driving force behind winds is still an open issue and we discuss only in passing how a simple iso-thermal wind model could fit our data. Nevertheless, the comparison with inferences from observations and theory looks promising with radio continuum offering a complementary view on galactic winds. We finish with a perspective on future observations and challenges lying ahead.
\end{abstract}


%
\section{Introduction}
\label{s:introduction}

Cosmic rays are one of the major ingredients in the interstellar medium (ISM), their energy density being comparable to that of the gaseous phases. Hence, cosmic rays play a major role in shaping the formation and evolution of galaxies in the Universe. The physics of cosmic rays is now investigated with multi-messenger astronomy \citep[see][for a recent review]{becker_tjus_20a}, with a focus on the Milky Way. In recent years, nearby galaxies have become accessible both with radio continuum \citep{irwin_12a} and $\gamma$-ray observations \citep{ackermann_12a} to better constrain cosmic-ray transport parameters. In this review, we present some observational inferences that have been made in the past few years with improved (i.e.\ more sensitive) radio continuum observations, and some of the advances made modelling them. Our aims are several-fold: first, we wish to explore the physics at cloud-scale at least in an indirect way, such as the entrainment of clouds in a hot wind \citep{brueggen_20a}. Second, the global structure of the ISM dynamics is studied -- something that can be well done for external galaxies -- and which may inform simulations from column-type simulations that can resolve the supernova blast waves on a 10-pc scale \citep{girichidis_18a} over global simulations of isolated galaxies \citep{salem_14a,Pakmor_16} to cosmological zoom-in simulations \citep{pakmor_17a}. Third, we can also explore the relationship with the magnetic field in the halo which fascinatingly takes the form of an X-shaped morphology \citep{tuellmann_00a,soida_11a} and compare this with models and simulations that include the effect of magnetic fields \citep{pakmor_17a,steinwandel_20a}. Our work may lead eventually to the necessary understanding, so that the frequently used simple recipes for `sub-grid physics' that are used in cosmological simulations of galaxy evolution to resemble observed galaxies \citep{vogelsberger_20a} to put on a sound physical basis.

We will in particular address the question to what extent cosmic rays can have an influence on galaxy evolution in the form of galactic winds \citep[see][for a recent review on the cold component of winds]{veilleux_20a}. Cosmic rays are thought to be responsible for winds that are `cooler and smoother' \citep{girichidis_18a} and so can lead to higher mass-loss rates than purely thermally driven winds. Also, cosmic ray-driven winds can be successful in environments that are more typical for $L_\star$ galaxies, such as our own Milky Way, and in particular our solar neighbourhood \citep{everett_08a}. These environments have much lower star-formation rate surface densities (\sfrd) with \sfrd $\approx 3\times 10^{-3}$~\usfrd, however, observationally they are more difficult to access than canonical star burst galaxies such as M~82 and the nuclear region in NGC~253. These `superwind' galaxies with \sfrd\ $\sim 10^{-1}$~\usfrd\ \citep{heckman_00a} are more extreme than the relatively benign late-type galaxies that have radio haloes \citep{wiegert_15a}. \citet{dahlem_95a} already suggested a low critical \sfrd-value based on radio continuum observations, which were later corroborated by optical emission line studies using integral field unit spectroscopy \citep{ho_16a,lopez_coba_19a}.

\begin{figure}[tb]
\centering
\includegraphics[width=0.8\columnwidth]{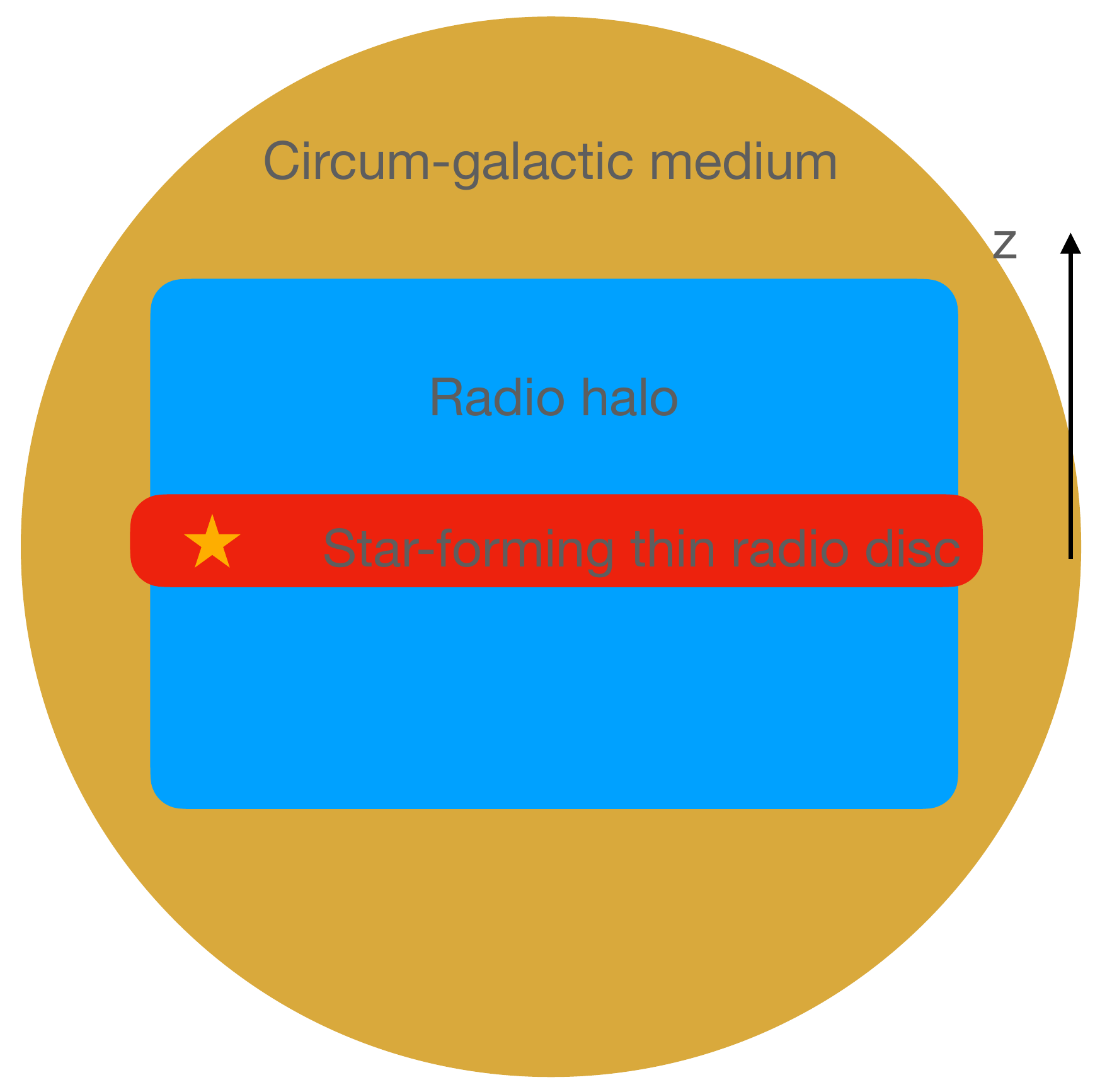}
\caption{The three principle components that we aim to study in the radio continuum of a galaxy as seen in the edge-on position} 
\label{fig:radiohalo}
\end{figure}

More generally speaking, we can explore which effects are driving galactic winds, with processes related to stellar feedback and active galactic nuclei (AGNs) the main candidates \citep{yu_20a}. Not only the mass-loss rates, but also the composition of the wind fluid is important for galaxy evolution as is the final fate of the gas and the relation that galaxies have with the circum-galactic medium \citep[CGM; see][for a recent review]{tumlinson_17a}. The main questions that we would like to address with the study of radio continuum haloes (see Fig.~\ref{fig:radiohalo}) are (i) how predominant are galactic winds?; (ii) what is the role of supernovae, radiation pressure, cosmic-ray pressure, and AGN? Is there a minimum threshold of star formation or black hole activity needed to trigger cool outflows?; (iii) what is the relative distribution of the cool, warm, and hot phases in the wind? (iv) What feedback effects do they exert on the host galaxy ISM and CGM?

%


Cosmic rays have become recently a hot candidate to drive galactic winds, although the basic idea was already explored by \citet{ipavich_75a}. Cosmic rays have a relatively soft equation of state mean that they build up a gentle pressure gradient in the halo with a scale height of $\sim$1~kpc. This pressure gradient can gently accelerate the gas, possibly in conjunction with the hot ionised gas \citep{breitschwerdt_93a,everett_08a,recchia_16a}. In order to build up the necessary pressure gradient, cosmic rays have to first diffuse out of the star-forming regions \citep{salem_14a}. This can be done by either diffusion or streaming \citep{uhlig_12a}; if the cosmic rays are only passively advected, they only act as an additional pressure component and so merely puff up the gaseous disc a bit more without leading to a wind \citep{farber_18a}. Besides creating a wind, cosmic rays may play a key role in accelerating clouds of cold gas via the `bottle neck effect' in which streaming plays an important role \citep{wiener_17a}, significantly boosting the mass-loss rate.


Radio continuum observations trace cosmic-ray electrons, the spectra of which give important clues on their transport. Early works on the integrated radio continuum spectra of galaxies showed that their curved spectra can be explained by a transition from escape-dominated radio haloes at low frequencies to radiation-loss dominated haloes at high frequencies \citep{pohl_91a}. The changing radio spectral index with distance from the star-forming mid-plane can be modelled with diffusion and advection, which result in different properties \citep{lisenfeld_00a}.

The analysis of the radio spectral index in external galaxies was for a long time limited by observations, where it is relatively hard to measure the radio spectral index of extended objects using radio interferometry, for instance by the limitations due to a lack of sufficiently short base lines. However, with new instruments such as the LOw-Frequency Array \citep[LOFAR;][]{vanHaarlem_13a}, the upgraded Jansky Very Large Array \citep[JVLA;][]{irwin_12a} and improved data reduction techniques, in particular image deconvolution with the multi-scale multi-frequency MS-MFS {\sc clean} algorithm \citep{rau_11a}, some of these limitations have now been overcome.



\subsection{A simplified overview of cosmic ray transport}
We follow the standard paradigm, where cosmic rays are accelerated and injected into the ISM at supernova remnants (SNRs) by diffusive shock acceleration \citep[DSA;][]{bell_78a}. On average, the kinetic energy per supernova is $10^{51}~\rm erg$, a few per cent of which is used for the acceleration of cosmic rays \citep[e.g.][]{rieger_13a}. The cosmic-ray luminosity of a galaxy is then \citep{socrates_08a}:
\begin{equation}
    L_{\rm CR} = 3(\epsilon_{\rm SN}/0.1)\rm (SFR / M_\odot\,yr^{-1}) \times 10^{40}~\rm erg\,s^{-1},
    \label{eq:cosmic_ray_luminosity}
\end{equation}
where $\epsilon_{\rm SN}$ is the energy conversion factor from SNe kinetic energy into cosmic rays. Of the energy stored in the cosmic rays, between 1 and 2 per cent is channelled into the cosmic-ray electrons with the rest into protons and heavier nuclei \citep{beck_05a}. 

Cosmic-ray transport proceeds either by diffusion along and across magnetic field lines, cosmic-ray streaming and advection \citep{ensslin_11a}.  Diffusion of cosmic rays can be understood as them being scattered at magnetic field irregularities and so following a stochastic path with a bulk speed much smaller than the speed of light. This view is corroborated by the fact that in the Milky Way the cosmic ray flux has a directional anisotropy of only $10^{-4}$ \citep{ahlers_17a}. Cosmic rays reside in the Galaxy for an energy-dependent time which is (1--$2)\times 10^{7}$~yr at 1 GeV and decreases as a low fractional power of energy \citep{zweibel_13a}. The turbulence of the magnetic fields can be either created by external processes such as supernovae and stellar winds that inject the turbulence at the tens of parsec scale, which cascades down to the cosmic-ray gyro radius; this case is usually referred to as cosmic-ray diffusion. Or cosmic rays can transfer some of their energy and momentum on the magnetic field thereby creating their own turbulence; this case is referred to as cosmic-ray streaming, where the cosmic rays follow the magnetic field lines too. 

The question which values of diffusion coefficients and streaming speeds to use is of importance for numerical simulations. Values for the diffusion coefficient range from $10^{27}$~\udif\ \citep{salem_14a} to more conventional values of $10^{28}$~\udif\ \citep{girichidis_18a} to even larger values of $10^{29}$--$10^{30}$~\udif\ \citep{hopkins_20a}. The canonical Milky Value of $3\times 10^{28}$~\udif\ \citep{strong_07a} is model-dependent, particularly on the size of the halo, so that the diffusion coefficient may potentially be higher if the halo is larger. In several works, a small diffusion coefficient is argued to be of importance so that the interaction with the gas is strong enough \citep{pakmor_16a}. In contrast, \citet{hopkins_20a} argue that the diffusion coefficient needs to be larger at $10^{29}$~\udif\ so that the $\gamma$-ray flux is not too high in star-forming galaxies. If anisotropic diffusion is modelled, the ratio of perpendicular to parallel diffusion coefficients is of importance but only poorly constrained with canonical values of $D_\perp/D_\parallel = 10$--100. Similarly, the velocity of cosmic-ray streaming is largely unknown although most theories agree that it should be of the order of the Alfv\'en speed. In the absence of ion-neutral damping, the wave growth of the Alfv\'en waves is unchecked so that cosmic rays can stream at super-Alfv\'enic speeds \citep{ruszkowski_17a}. 

\subsection{Review structure}
\label{ss:review_structure}

A study of cosmic ray transport in external galaxies aims to determine the value of the diffusion coefficient including its energy dependence, whether diffusion proceeds isotropic or anisotropic and to what extent streaming takes over diffusion in galactic discs as the dominant transport process. In order to do this we exploit synchrotron emission from cosmic-ray electrons. As cosmic rays are injected at star formation sites, the smearing out of the radio continuum emission with respect to the star-formation distribution allows us to measure the cosmic-ray transport length. In conjunction with spectral ageing, we can model cosmic-ray transport using the electrons as proxies. This is the basic idea of our approach. 

This review is structured as follows. In Section~\ref{s:methodology}, we introduce the methodology used in order to interpret the radio continuum observations. Section~\ref{s:spinnaker} gives an overview of the software {\sc spinnaker}, which we have developed to model the observations. The next three sections provide an overview on the different methods that have been used: in Section~\ref{s:radio_haloes}, we present our inferences that we can gain from the vertical intensity profiles in edge-on galaxies; Section~\ref{s:radio_continuum_spectrum} summarises what we can learn from the radio continuum spectrum; in  Section~\ref{s:face_on_galaxies}, we extend this approach to face-on galaxies. In Section~\ref{s:results}, we summarise the most important results from our studies thus far. These results motivate a new approach to model radio haloes by stellar feedback-driven winds as laid out in Section~\ref{s:wind}. We put our results into context of inferences from absorption- and emission-line studies in Section~\ref{s:optical_inferences} and to inferences from theory in Section~\ref{s:inferences_from_theory}. In Section~\ref{s:missing_physics}, we discuss missing physics from our models thus far and how to address this shortcoming in the future. In Section~\ref{s:summary}, we summarise.

\section{Methodology}
\label{s:methodology}

\subsection{Radio continuum emission from galaxies}
Radio continuum emission from galaxies traces cosmic-ray electrons (CR$e^{-}$), emitting synchrotron emission while spiralling around magnetic field lines. The other contribution is from \emph{thermal} emission, which stems from the free--free emission of thermal electrons; for this contribution, the thermal H\,$\alpha$ emission is a good tracer and so that the emission can be separated if desired.

In the interstellar medium, CR$e^{-}$ are losing their energy mainly due to
synchrotron and inverse Compton (IC) radiation, so that GeV-electrons have lifetimes of a few $10^7~\rm
yr$. The ionization and bremsstrahlung losses for typical ISM
densities of $n=0.05~\rm cm^{-3}$ result in lifetimes of the order of $10^9~\rm yr$
and can hence be neglected \citep{heesen_09a}, except at low frequencies in dense gaseous, star-forming regions \citep{basu_15a}. A comparison of $\gamma$-ray luminosity with Monte--Carlo simulations have shown that cosmic rays sample the mean density of the interstellar medium \citep{boettcher_13a}, hence such an assumption may be justified. The combined synchrotron and IC loss rate for CR$e^{-}$ is given by \citep{longair_11a}:
\begin{equation}
-\left (\frac{{\rm d}E}{{\rm d}t}\right )=b(E)=\frac{4}{3} \sigma_{\rm T} c \left (\frac{E}{m_{\rm
      e}c^2} \right )^2 (U_{\rm rad}+U_{\rm B}),
\label{eq:be}
\end{equation}
where $U_{\rm rad}$ is the radiation energy density, $U_{\rm B}=B^2/8\pi$ is the magnetic
 energy density, $\sigma_{\rm T}=6.65\times 10^{-25}~\rm cm^2$ is the
Thomson cross-section and $m_{\rm e}=511~\rm keV\,c^{-2}$ is the electron
rest mass. The CR$e^{-}$ energy can be inferred from the critical frequency, where the synchrotron spectrum peaks for an individual electron \citep{beck_15a}:
\begin{equation}
    E \approx \left(\frac{\nu}{16~\rm MHz}\right )^{1/2} \left(\frac{\mu\rm G}{B_{\perp}}\right)^{1/2},
    \label{eq:cre_energy}
\end{equation}
where $B_{\perp}$ is the total magnetic field strength perpendicular to the line of sight (i.e. in the sky plane). The time dependence of the energy for an individual CR$e^{-}$ is $E(t)=E_0(1+t/t_{\rm syn})^{-1}$,
so that at $t=t_{\rm syn}$ the energy has dropped to half of its initial
energy $E_0$. The CR$e^{-}$ synchrotron lifetime, as determined by synchrotron losses, and a smaller contribution from IC radiation losses, can be expressed by \citep{heesen_16a}:
\begin{eqnarray}
t_{\rm syn} & = &34.2 \left (\frac{\nu}{\rm 1\,GHz}\right )^{-0.5}
\left (\frac{B}{\rm 10\,\mu G}\right )^{-1.5} \\\nonumber & & \left
  (1+\frac{U_{\rm rad}}{U_{\rm B}}\right )^{-1}~{\rm Myr}.
\label{eq:t_syn}
\end{eqnarray}

If the CR$e^{-}$ escape time is $t_{\rm esc}$, the effective CR$e^{-}$ lifetime is then:
\begin{equation}
\tau^{-1} = t_{\rm syn}^{-1} + t_{\rm esc}^{-1}.
\label{eq:cre_lifetime}
\end{equation}

The CR$e^{-}$ injection spectrum ${\rm d}EN(E) = N_0 E^{-\gamma_{\rm inj}}$ is a power-law with an injection spectral index of $\gamma_{\rm inj}\approx 2.2$ \citep[fig.~3a in][]{caprioli_11a}. Hence, the integrated radio continuum spectrum can give us important clues about the escape of CR$e^{-}$ because, depending on the energy dependence of the various loss processes, the injection spectrum is converted into a power-law with a different slope. For instance, the spectrum is steepened to $\propto E^{-\gamma_{\rm inj}-1}$ if the energy losses are proportional to $E^2$ as is the case for both synchrotron and IC radiation losses \citep{longair_11a}. This means that the radio spectral index is steepened to $\alpha=\alpha_{\rm inj} - 0.5$, where $\alpha_{\rm inj}=(1-\gamma_{\rm inj})/2$ is the injection radio spectral index.\footnote{Radio spectral indices are defined as $I_\nu\propto \nu^{\alpha}$.} Thus, in galaxies with free CR$e^{-}$ escape, the radio continuum spectrum is a power-law with $\alpha\approx -0.6$. Contrary, if the CR$e^{-}$ losses due to synchrotron and IC losses are important, the spectrum steepens to $\alpha\approx -1.2$ \citep{lisenfeld_00a}.

\subsection{Advection--diffusion approximation}

The CR$e^{-}$ energy spectrum $N(E){\rm d}E$ can be
modelled by solving
the diffusion--loss equation for the CR$e^{-}$ \citep[e.g.][]{longair_11a}:
\begin{equation}
  \frac{{\rm d} N(E)}{{\rm d}t} = D \nabla^2 N(E) +
  \frac{\partial}{\partial E}\left[ b(E) N(E)\right ] + Q(E,t),
\label{eq:diffloss}
\end{equation}
where $b(E)=-{\rm d}E/{\rm d}t$ for a single CR$e^{-}$ as given by Equation~(\ref{eq:be}). Massive
spiral galaxies have rather constant star formation histories, so that the CR$e^{-}$ injection
rate can be assumed as approximately constant and so the source term $Q(E,t)$ has no explicit time dependence. If we assume that all sources of CR$e^{-}$ are located in the disc plane, we obtain for the source term $Q(E,t)=0$ for
$z>0$ (Fig.~\ref{fig:radiohalo}). Equation~(\ref{eq:diffloss}) can be evolved in time until a stationary solution is found. We use a slightly different approach, first by restricting ourselves to a one-dimensional (1D) problem, and second by imposing a fixed inner boundary condition of $N(E, 0) = N E^{-\gamma_{\rm inj}}$. In the stationary case, the change of the CR$e^{-}$ number density $\partial N/\partial t$ is solely determined by the energy loss term (second term on the right-hand side of equation~\ref{eq:diffloss}). Noticing that for advection we have $\partial N /\partial t = v \partial N /\partial z$, we can re-write equation~\eqref{eq:diffloss} for the case of pure advection to:

\begin{equation}
    \frac{\partial N}{\partial z} = \frac{1}{v} \left \lbrace \frac{\partial}{\partial E} [b(E)N(E,z)]\right \rbrace, 
    \label{eq:n_advection}
\end{equation}
where $v$ is the advection speed, assumed here to be constant. Similarly, for diffusion we have $\partial N /\partial t = D\partial N^2 /\partial z^2$ (Fick’s second law of diffusion), so that we can re-write equation~\eqref{eq:diffloss} for the case of pure diffusion to:
\begin{equation}
    \frac{\partial^2 N}{\partial z^2} = \frac{1}{D} \left \lbrace \frac{\partial}{\partial E} [b(E)N(E,z)]\right \rbrace ,
    \label{eq:n_diffusion}
\end{equation}
where the diffusion coefficient can be parametrised as function of energy as $D = D_0(E/{\rm GeV})^{\mu}$. If the diffusion coefficient is energy-dependent, values for $\mu$ are thought to be between $0.3$ and $0.6$ \citep*{strong_07a}. For diffusion we also assume that the halo size is much larger than the CR$e^{-}$ diffusion length and so the CR$e^{-}$ cannot escape at the halo boundary, and the decrease of the CR$e^{-}$ number density is solely determined by the energy losses (synchrotron and IC radiation).

If we drop the assumption of a constant advection speed, the CR$e^{-}$ number density will change even if the cross-sectional area $A$ of the outflow is constant. According to the continuity equation:
\begin{equation}
    n_{\rm CR} v A = \rm const.,
\end{equation}
where $v$ is the advection speed and $n_{\rm CR}$ is CR$e^{-}$ number density. Additionally, there are adiabatic losses (cooling) that can be described as:
\begin{equation}
    -\left (\frac{{\rm d}E}{{\rm d}t}\right ) = \frac{1}{3}(\nabla\cdot v)E = \frac {E}{t_{\rm ad}}.
\end{equation}{}
For a linearly accelerating wind with a constant cross-sectional area, the adiabatic loss time-scale is:
\begin{equation}
    t_{\rm ad} = 3\left (\frac{{\rm d}v}{{\rm d}z}\right)^{-1}.
\end{equation}
An outflow that is either expanding laterally with an increasing cross-section or accelerating hence leads to adiabatic losses. Both effects can of course also work in combination, which decrease the cosmic-ray energy density, such that the cosmic rays can be in equipartition with the magnetic field. Assuming that the cosmic rays are in equipartition with the magnetic field in the disc plane, a constant advection speed in conjunction with a non-expanding outflow leads to a severe violation of equipartition in the halo \citep{mora_19a}.

We also have to assume a magnetic field distribution. Because of simplicity we first parametrise the magnetic field as exponential distribution, so that the magnetic field strength is:
\begin{equation}
    B(z) = B_0 \exp(-z/h_{\rm B}),
    \label{eq:one_comp_bfield}
\end{equation}
where $h_{\rm B}$ is the magnetic field scale height. The magnetic field strength in the mid-plane $B_0$ is then a fixed parameter calculated with the revised equipartition formula \citep{beck_05a}. Alternatively, we also use a two-component exponential magnetic field:
\begin{equation}
    B(z) = B_1 \exp(-z/h_{\rm B1}) + B_2 \exp(-z/h_{\rm B2}),
    \label{eq:two_comp_bfield}
\end{equation}
where $h_{\rm B1}$ and $h_{\rm B2}$ are the magnetic field scale heights in the thin and thick radio disc, respectively, with the magnetic field strengths related as $B_0=B_1+B_2$. The thick radio disc is also referred to as radio halo (see Fig.~\ref{fig:radiohalo}).
%
\begin{figure}[tb]
\includegraphics[width=\columnwidth]{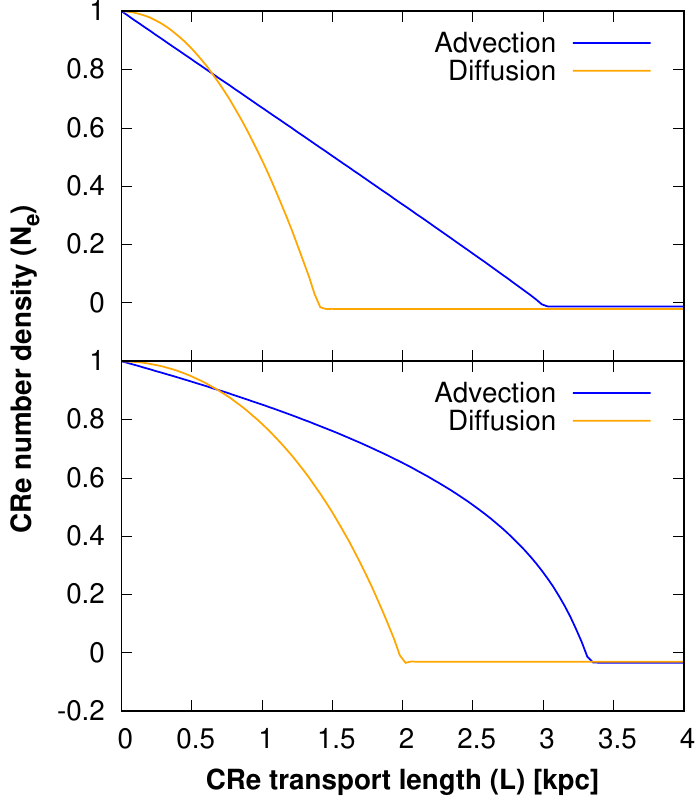}
\caption{Comparison of advection and diffusion, where we plot the normalised CR$e^{-}$ number density as function of the CR$e^{-}$ transport length. The CR$e^{-}$ injection spectral index is $\gamma_{\rm inj}=3$ (\emph{top panel}) and $\gamma_{\rm inj}=2$ (\emph{bottom panel}). Further parameters as described in the text are $B=10~\mu\rm G$, $U_{\rm rad}/U_{\rm B}=0.3$, $D=10^{28}~\rm cm^2\,s^{-1}$, and $v=100~\rm km\, s^{-1}$. Models adopted from \citet{heesen_16a}} 
\label{fig:t_syn}
\end{figure}

\subsubsection{Cosmic-ray electron transport length}
\label{sss:cre_transport_length}
With the most simplistic description, the cosmic-ray diffusion length can be described as:
\begin{equation}
    D = \frac{L^2}{4\tau},
    \label{eq:diffusion}
\end{equation}{}
where $D$ is the isotropic diffusion coefficient and $\tau$ is the CR$e^{-}$ lifetime. Hence, it follows that the cosmic-ray transport length $L$ scales only with the square root of the CR$e^{-}$ lifetime as $L=\sqrt{4D\tau}$. Using convenient units, we find:
\begin{equation}
    D = 0.75\times 10^{29}~\frac{(L/{\rm kpc})^2}{\tau/{\rm Myr}}~\rm cm^2\,s^{-1}.
    \label{eq:diffusion_coefficient}
\end{equation}{}
Conversely, advection can be simply described as:
\begin{equation}
    v = \frac{L}{\tau},
    \label{eq:advection}
\end{equation}{}
where $v$ is the advection speed. Or, in convenient units:
\begin{equation}
    v = 980\frac{L/{\rm kpc}}{\tau/{\rm Myr}}~\rm km\,s^{-1}.
\end{equation}
For advection, the CR$e^{-}$ transport length scales linearly with the CR$e^{-}$ lifetime as $L=v\tau$. For small CR$e^{-}$ lifetimes, diffusion happens faster than advection and so diffusion dominates over advection near the sources in the star-forming disc. Equating diffusion and advection length, $\sqrt{4D\tau}=v\tau$, the CR$e^{-}$ lifetime becomes:
\begin{equation}
    \tau = \frac{4D}{v^2},
\end{equation}{}
or, in convenient units:
\begin{equation}
    \tau = 12\frac {D / 10^{28}~{\rm cm^2\,s^{-1}}}{(v/100~{\rm km\,s^{-1})^2}}~\rm Myr.
\end{equation}{}
Inserting this lifetime into equation~\eqref{eq:advection}, we obtain the cosmic-ray transport length, where the transition from diffusion to advection happens:
\begin{equation}
    z_{\star} \approx 1.2 \frac{D / 10^{28}~{\rm cm^2\,s^{-1}}}{v / {\rm 100~\rm km\,s^{-1}}}~\rm kpc.
    \label{eq:diff_adv}
\end{equation}{}

%
\begin{figure*}[tb]
\includegraphics[width=\textwidth]{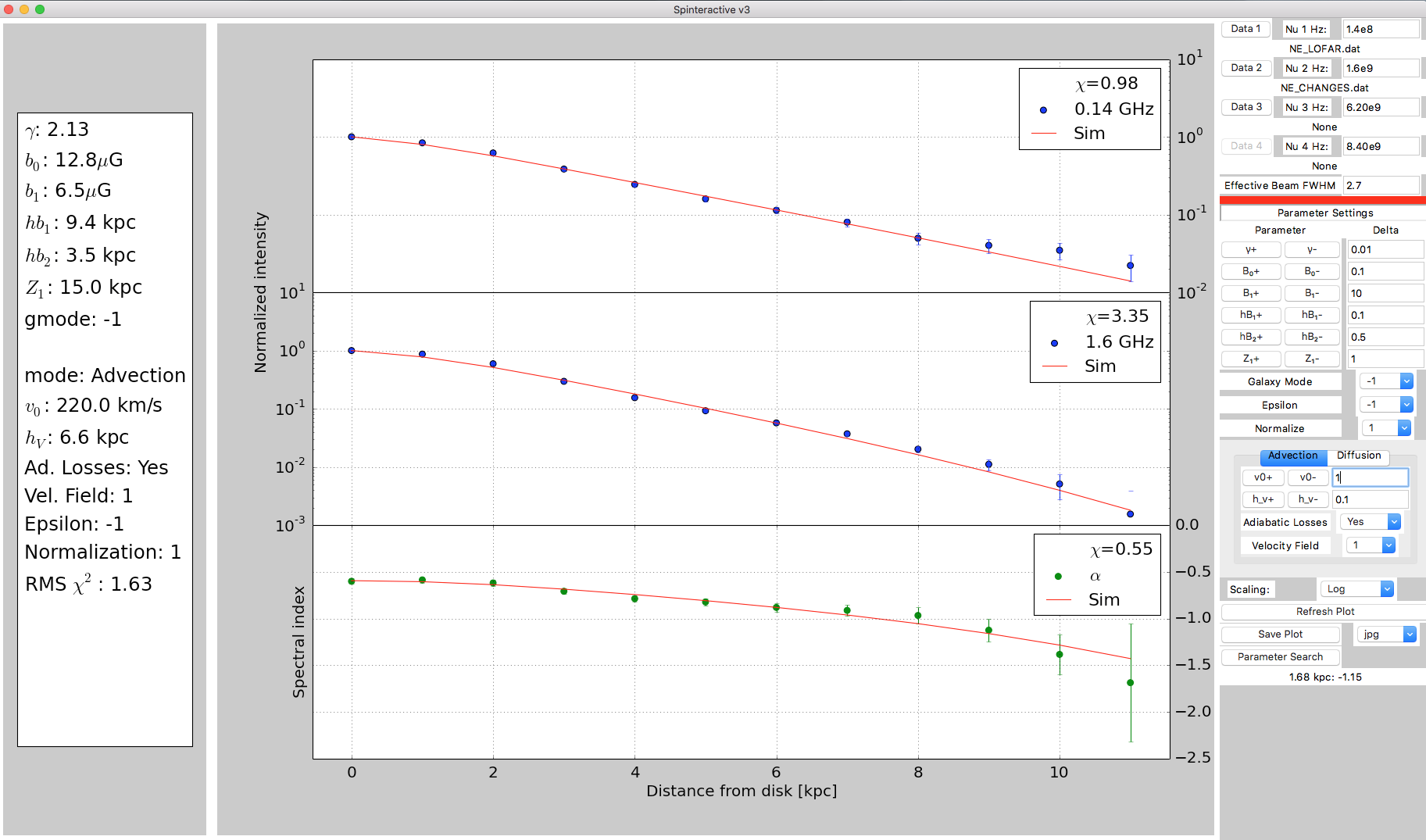}
\caption{{\sc spinnaker} as viewed with {\sc spinteractive}. Application in NGC~5775 to LOFAR 150-MHz and CHANG-ES 1.5-GHz data. In the left panel, the parameters are summarised. The middle panel shows from top to bottom, the 140 MHz data, the 1.5-GHz data, and the radio spectral index. The right panel shows the parameters than can be interactively changed.} 
\label{fig:spinteractive}
\end{figure*}

The diffusion-dominated region near the mid-plane extends to heights of $z\lesssim z_\star$, whereas the advection-dominated region in the halo is at heights of $z\gtrsim z_\star$ \citep{recchia_16a}.
In Fig.~\ref{fig:t_syn}, we plot the CR$e^{-}$ number density both for advection and diffusion as function of the CR$e^{-}$ transport length. The transition happens at about $0.6$~kpc, where for diffusion the CR$e^{-}$ number density drops rapidly and so advection takes over as the dominating transport mode. For the modelling of the cosmic-ray transport it is hence useful to approximate the transport by \emph{pure advection} if the galaxy has a wind because diffusion is suppressed in the halo, where we model the data. In contrast, if a galaxy has no wind, we can approximate the transport by \emph{pure diffusion}. This is the approach we take in the following.

\subsection{Expected relations}
\label{ss:expected_relations}
Intensity scale heights in edge-on galaxies can be used in two ways in order to investigate the cosmic-ray transport. For both methods, we use the \emph{equipartition} assumption to derive the CR$e^{-}$ scale height from the non-thermal intensity scale height by:
\begin{equation}
    h_e = \frac{3 -\alpha_{\rm nt}}{2} h_{\rm syn}.
    \label{eq:h_e}
\end{equation}{}
The first method is then to measure the scale height at two different frequencies (or more), where the different frequency-dependence of the scale height can be used to distinguish between advection and diffusion. Combining the CR$e^{-}$ synchrotron lifetime (equation~\ref{eq:t_syn}) with the advection transport length (equation~\ref{eq:advection}), we obtain for \emph{advection}:
\begin{equation}
    h_{\rm e} \propto \nu^{-0.5}B^{-3/2}.
    \label{eq:advection_scale_height}
\end{equation}{}
Similarly, using the diffusion transport length (equation~\ref{eq:diffusion}), we obtain for \emph{diffusion}:
\begin{equation}
    h_{\rm e} \propto \nu^{(\mu-1)/4}B^{-(\mu+3)/4}.
    \label{eq:diffusion_scale_height}
\end{equation}{}

Hence for diffusion, the CR$e^{-}$ scale height depends less on the frequency than for advection. For a possible energy-dependence of the diffusion coefficient, this  frequency dependence of the CR$e^{-}$ scale height is reduced even further such that for a hypothetical, strong energy-dependence of the diffusion coefficient with $\mu=1$, the frequency-dependence of the scale height even vanishes entirely. 

It is important to be aware of that above relations only apply as long as the energy losses of the CR$e^{-}$ are high, as is for instance the case if the magnetic field strength in the halo is constant and so the CR$e^{-}$ lose all their energy. This scenario is referred to as the \emph{calorimetric case}. More realistically, galaxies may lose some of their CR$e^{-}$ or the CR$e^{-}$ even escape almost freely from the galaxy, referred to as \emph{non-calorimetric} case. For the latter, we do not expect any dependence of the scale height on frequency. This is the case if the escape time-scale:
\begin{equation}
    t_{\rm esc} = \frac{h_{\rm e}}{v}
\end{equation}
is much smaller than the CR$e^{-}$ lifetime, i.e. $t_{\rm esc}\ll \tau$.

Since the CR$e^{-}$ lifetime depends most on the frequency and the magnetic field strength, attempts so far have concentrated on measuring the CR$e^{-}$ transport length as function of them. Even more challenging is to quantify the influence of the magnetic field structure, the influence of which on the anisotropic parallel diffusion coefficient can be parametrised as \citep{shalchi_09a}:
\begin{equation}
    D_\parallel \propto \left (\frac{B_{\rm ord}}{B_{\rm turb}}\right )^2B_{\rm ord}^{-1/3},
\end{equation}
where $B_{\rm ord}$ is the ordered magnetic field strength and $B_{\rm turb}$ is the turbulent magnetic field strength. As we shall see, the main challenge is in separating the effects of spectral ageing and the influence of the magnetic field. The basic idea is to use the equations for advection and diffusion to separate them. In order to do this we implemented them in a simple-to-use computer program.

%
\begin{table}[tb]
\small
\caption{Parameters in {\sc spinnaker}} 
 \label{tbl:par}
\begin{tabular}{lc}
 \tableline  
  Parameter & fitted \\
  \tableline  
  Magnetic field strength $B_0$ & fixed \\
  Injection CRe spectral index $\gamma_{\rm inj}$ & fitted \\
  \multicolumn{2}{c}{-- Diffusion: mode = 1 --} \\
  Diffusion coefficient $D$ & fitted \\
  Energy dependence $\mu$ & fitted \\
  \multicolumn{2}{c}{-- Advection: mode = 2 --} \\
  \multicolumn{2}{c}{Constant speed ($\tt velocity\_field = 0$)} \\ 
  Advection speed $v_0$ & fitted \\
 \tablenotemark{a} Magnetic field scale height $h_{\rm B}$ & fitted \\
  \multicolumn{2}{c}{Exponential velocity profile ($\tt velocity\_field = 1$)} \\
  Advection speed (at $z = 0$) $v_0$ & fitted \\
  \tablenotemark{a} Magnetic field scale height $h_{\rm B}$ & fitted \\
  Velocity scale height $h_v$ & fitted \\
  \multicolumn{2}{c}{Power-law velocity profile ($\tt velocity\_field = 2$)} \\
  Velocity scale height $h_v$ & fitted \\
  Advection speed (at $z=0$) $v_0$ & fitted \\
  \tablenotemark{a} Magnetic field scale height $h_{\rm B}$ & fitted \\
  Velocity scale height $h_v$ & fitted \\
  Power-law index $\beta$ & fitted \\
  \multicolumn{2}{c}{Wind model ($\tt velocity\_field = 3$)}\\
  Advection speed (at $z=z_{\rm c}$) $v_0$ & fitted \\
  Flux tube opening power $\beta$ & fitted \\
  Flux tube scale height $z_0$ & fitted \\
  \tableline 
 \end{tabular}
%
 \tablenotetext{a}{In case of a 1-component exponential magnetic field;\\ there is also the option to fit a 2-component exponential\\ magnetic field ($\tt galaxy\_mode=1$).}
 \end{table}

%
\begin{figure}[tb]
\includegraphics[width=\columnwidth]{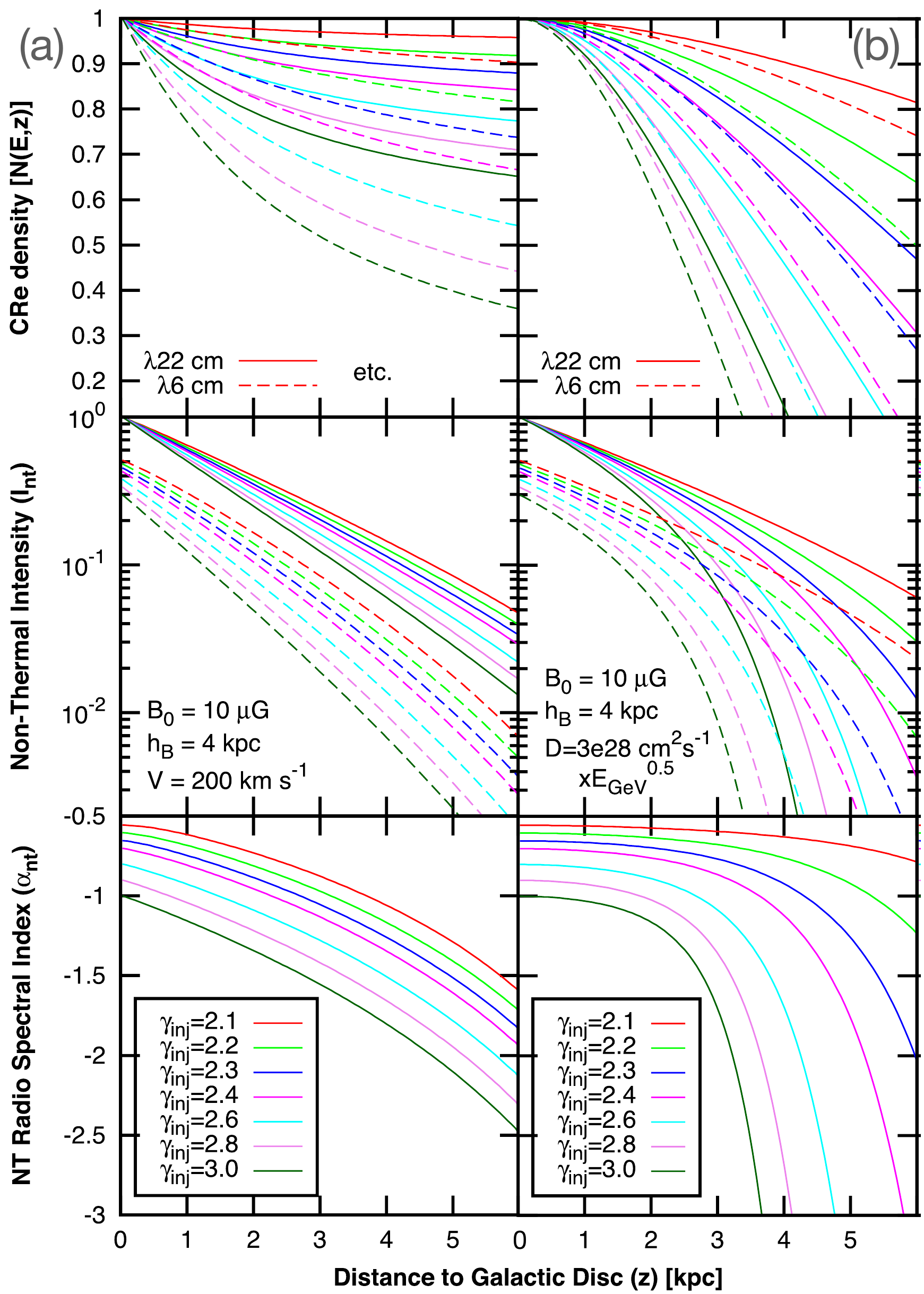}
\caption{Family of {\sc spinnaker} models for various CR$e^{-}$ injection spectral indices at 1.4~GHz (solid lines) and 5~GHz (dashed lines). (a) is for advection and (b) is for diffusion. The magnetic field is a 1-component exponential function with $B=10~\mu\rm G\exp(-z/4~{\rm kpc})$, the advection speed is constant with $v_0=200~\rm km\,s^{-1}$ and the diffusion coefficient is $D=3\times 10^{28}(E/{\rm GeV})^{0.5}$~\udif. The first row shows the CR$e^{-}$ number density, the second row the non-thermal intensity and the third row the non-thermal radio spectral index between $1.4$ and 5~GHz. From \citet{heesen_16a}} 
\label{fig:diffadv}
\end{figure}

\section{An overview of {\sc spinnaker}}
\label{s:spinnaker}
The above equations were implemented in the computer program SPectral INdex Numerical Anlysis of K(c)osmic-ray Electron Radio-emission ({\sc spinnaker}).\footnote{https://github.com/vheesen/Spinnaker} The interactive version {\sc spinteractive} allows one the fitting of the intensities and radio spectral index profiles in a convenient way (see Fig.~\ref{fig:spinteractive}). In Table~\ref{tbl:par} we present the parameters that are fitted in each model. We now present the various options. 

Before we present the various options, we  briefly summarise the degeneracies involved in the empirical modelling, in particular with respect to the CR$e^{-}$ density, advection velocity, and diffusion coefficient. We assume the magnetic field strength in the disc as a fixed parameter, to be measured from the energy equipartition between cosmic rays and magnetic field. The main degeneracy we have to resolve is that either a high advection speed or diffusion coefficient will lead to a higher CR$e^{-}$ density in the halo, which can compensate a weaker magnetic field such that it still matches the observed level of intensity. Conversely, a strong magnetic field can compensate a lower CR$e^{-}$ density in the halo resulting in the same radio continuum intensity. This degeneracy can be be resolved by using the radio spectral index, since a higher advection speed or diffusion coefficient will lead to a flatter radio spectral index profile as the ageing of CR$e^{-}$ is suppressed. This is the reason why this kind of modelling can work at all and so we get fairly reliable values for either the diffusion coefficient and/or the advection speed \citep{heesen_16a}.

\subsection{Diffusion}
Pure diffusion is chosen by ${\tt mode=1}$, where the diffusion equation~\eqref{eq:n_diffusion} provides us with the CR$e^{-}$ number density profile as presented in Fig.~\ref{fig:diffadv}(b).  As can be seen, the diffusion approximation results in flatter CR$e^{-}$ number density profiles in the inner parts of the galaxy but steeper in the outskirts. The corresponding radio intensity profiles can be thus better described as Gaussian rather than as exponential functions \citep{heesen_19a}. The models presented in Fig.~\ref{fig:diffadv}(b) assume a non-constant, exponential magnetic field; while the magnetic field distribution influence the intensity profiles, the profiles are still markedly different from those as for advection (see also Fig.~\ref{fig:t_syn}). We also note that the profiles of the radio spectral index are also affected by this and have a `parabolic' shape. For diffusion we fit both the diffusion coefficient and the energy dependency $\mu$. 


\subsection{Advection}

The option $\tt mode = 2$ selects pure advection for the CR$e^{-}$ transport, where the CR$e^{-}$ number density is calculated according to equation~\eqref{eq:n_advection}.

\subsubsection{Constant advection speed}

For $\tt velocity\_field=0$, the advection speed is constant, which means the CR$e^{-}$ number density is regulated by radiation losses only. Hence, the CR$e^{-}$ number density decreases gradually with distance, different to diffusion (see Fig.~\ref{fig:diffadv}a). The radio spectral index is then also more gradually steepening in contrast to the diffusion solution, so that a linear function is a better fit.

\begin{figure}[tb]
\includegraphics[width=\columnwidth]{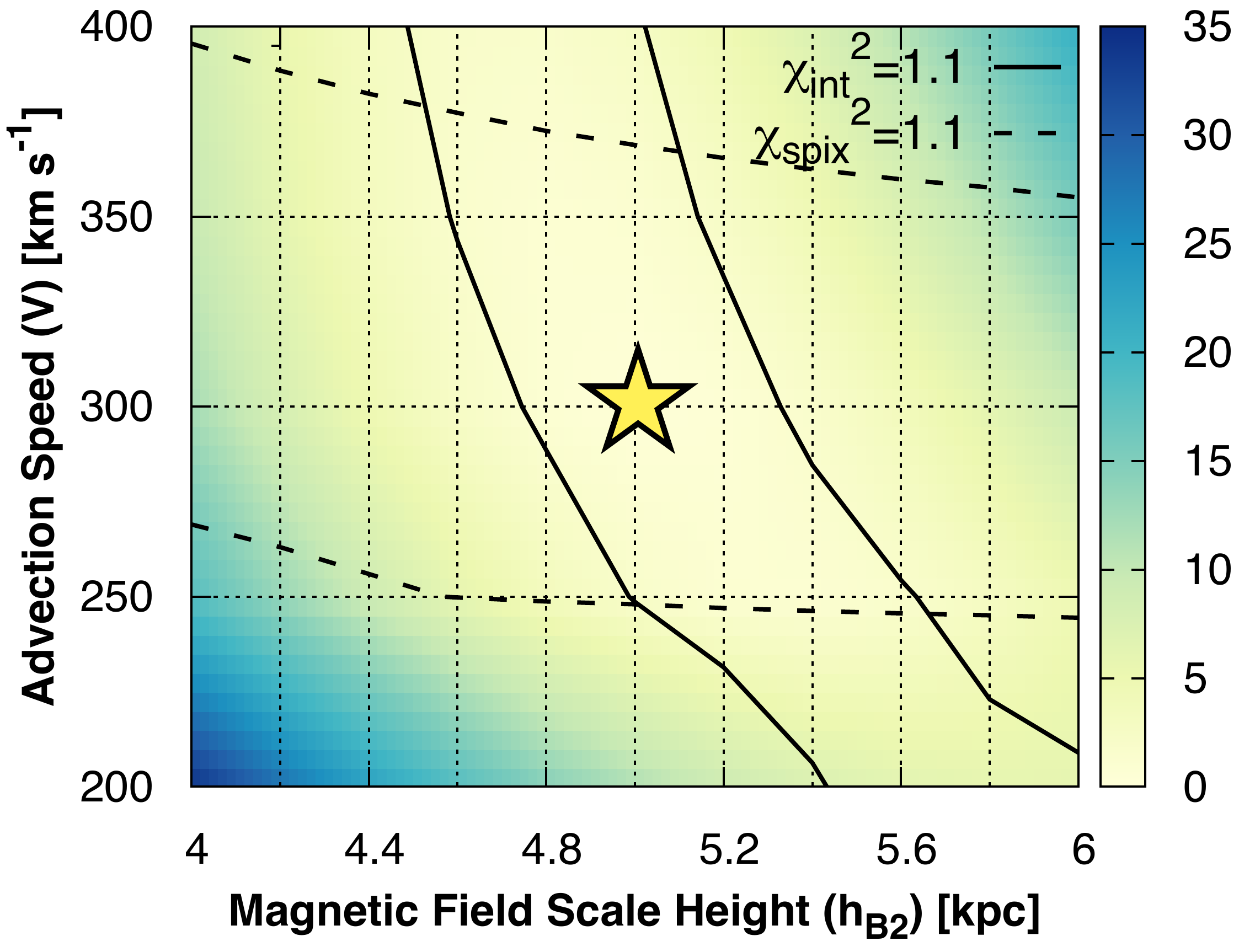}
\caption{Reduced $\chi^2$ for advection with a constant speed in the northern halo of NGC~4631. Solid lines (best-fitting intensities) define a diagonal area running from the top-left to the bottom-right, so that there is a degeneracy between advection speed and magnetic field scale height. The solution is unique though because the radio spectral index requires advection velocities as indicated by dashed lines, almost independent on the magnetic field scale height. The overlapping area in the  middle with the star in the centre, define the allowed best-fitting solutions. From \citet{heesen_18b}} 
\label{fig:chi2}
\end{figure}

For advection with a constant wind speed, we fit simultaneously for the advection speed $v_0$ and the magnetic field scale height. In principle, there is a \emph{degeneracy} between the advection speed and the magnetic field scale height if only one of the intensities are studied: a smaller magnetic field scale height can be compensated by a larger advection speed. However, the radio spectral index is also very dependent on the advection speed and so a unique solution can be found (Fig.~\ref{fig:chi2}). Depending on whether the vertical profile needs one or two magnetic field components (equations~\eqref{eq:one_comp_bfield} and \eqref{eq:two_comp_bfield}), we also may need to fit the magnetic field strength $B_1$ and scale height $h_{\rm B1}$ of the thin radio disc. If the angular resolution is sufficiently high to resolve the thin disc, it may be beneficial to only fit the radio spectral index in the halo, where advection dominates \citep{heesen_18b}.

%

The assumption of a constant advection speed has the advantage that the advection speeds can be accurately measured and these speeds can be regarded as a lower limits. The downside is that the cosmic-ray energy density is not in equipartition with the magnetic field for which an accelerating wind is necessary \citep{mora_19a}.

%
\begin{figure*}[tb]
\centering
\includegraphics[width=\textwidth]{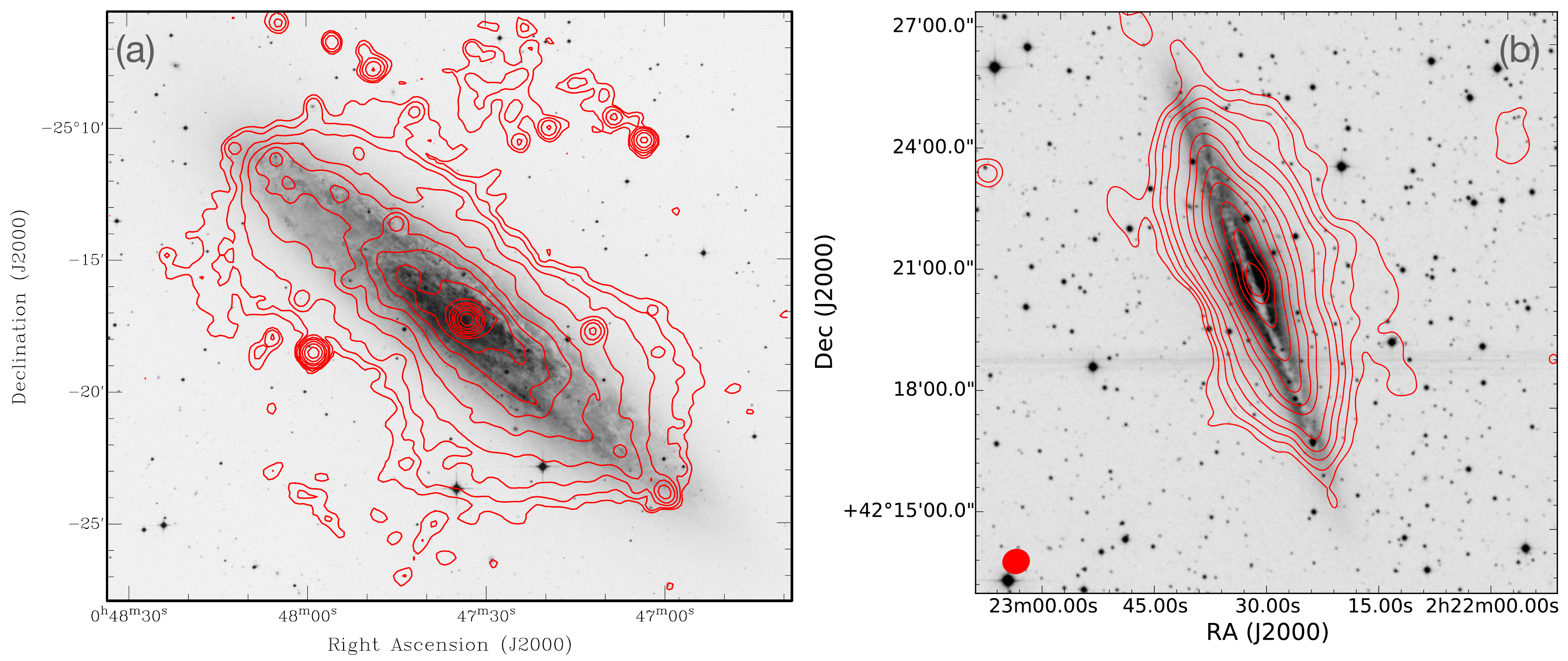}
\caption{Examples for two prominent dumbbell-shaped radio haloes with NGC~253 (a) and NGC~891 (b). Shown is the radio continuum intensity at 4850~MHz (a) and 146~MHz (b) as contours overlaid on optical images from the Digitized Sky Survey (DSS). From \citet{heesen_09a} (a) and \citet{mulcahy_18a} (b), respectively} 
\label{fig:dumbbell}
\end{figure*}

%
\begin{table*}[tb]
\small
\caption{Galaxies so far analysed with {\sc spinnaker} and re-evaluated for this review} 
 \label{tbl:sample}
\begin{tabular}{lcccccccc}
 \tableline  

 Galaxy & $d$\tablenotemark{a} & SFR\tablenotemark{b} &  $\log_{10}(\Sigma_{\rm SFR})$\tablenotemark{c}  &  $B_0$\tablenotemark{d} & $v_{\rm rot}$\tablenotemark{e} &
 Trans\tablenotemark{f} & $v$\tablenotemark{g} & Reference\\
 & (Mpc) & ($\rm M_\odot\,yr^{-1}$) & ($\rm M_\odot\,yr^{-1}\,kpc^{-2}$) & ($\mu\rm G$) &
 ($\rm km\,s^{-1}$) & & ($\rm km\,s^{-1}$) \\
  \tableline  
 IC 10   & 0.8  & 0.05  & $-1.00$ & 12.0 & 36  & Adv  & $29\pm 8$  & \citet{heesen_18c}\\
NGC 55   & 1.9  & 0.16  & $-2.87$ & 7.9  & 91  & Adv  & $100\pm 50$ & \citet{heesen_18b}\\
NGC 253  & 3.9  & 6.98  & $-1.80$ & 14.0 & 205 & Adv  & $400\pm 100$ & \citet{heesen_18b}\\
NGC 891  & 10.2 & 4.72  & $-1.96$ & 14.7 & 212 & Adv  & $150\pm 50$ & \citet{schmidt_19a}\\
NGC 3044 & 4.4  & 2.03  & $-2.17$ & 13.1 & 153 & Adv  & $200\pm 130$ & \citet{heesen_18b}\\
NGC 3079 & 7.7  & 9.09  & $-1.79$ & 19.9 & 208 & Adv  & $350\pm 70$ & \citet{heesen_18b}\\
NGC 3556 & 14.09 & 4.31 & $-2.26$ & $9.0$ & 154 & Adv & $145\pm 30$ & \citet{miskolczi_19a} \\
NGC 3628 & 14.8 & 1.73  & $-2.32$ & 12.6 & 215 & Adv  & $250\pm 150$ & \citet{heesen_18b}\\
NGC 4013 & 16.0  & 0.5   & $-2.61$ & 6.6  & 195 & Diff & $20\pm 10$  & \citet{stein_19b}\\
NGC 4217 & 20.6 & 4.61 & $-2.44$ & 11.0 & 195 & Adv & $350\pm 100$ & \citet{stein_20a} \\
NGC 4565 & 11.9 & 0.73  & $-2.88$ & 6.0  & 244 & Diff & $60\pm 30$  & \citet{heesen_19b}\\
NGC 4631 & 6.9  & 2.89  & $-2.20$ & 13.5 & 138 & Adv  & $250\pm 60$ & \citet{heesen_18b}\\
NGC 4666 & 26.6 & 16.19 & $-1.58$ & 18.2 & 193 & Adv  & $310\pm 50$ & \citet{stein_19a}\\
NGC 5775 & 26.9 & 9.98  & $-1.81$ & 16.3 & 187 & Adv  & $400\pm 80$ & \citet{heesen_18b}\\
NGC 7090 & 10.6 & 0.62  & $-2.41$ & 9.8  & 124 & Adv  & $200\pm 160$ & \citet{heesen_18b}\\
NGC 7462 & 13.6 & 0.28  & $-2.80$ & 9.7  & 112 & Diff & $90\pm 30$  & \citet{heesen_18b}\\

  \tableline 
 \end{tabular}
%
 \tablenotetext{a}{Assumed distance to galaxies;}
 \tablenotetext{b}{Star-formation rate (SFR), calculated from either total or mid-infrared luminosity;}
 \tablenotetext{c}{SFR surface density defined as \sfrd $=SFR/(\pi r_{\star}^2)$, where $r_{\star}$ is radius of the star-forming disc;} 
 \tablenotetext{d}{Magnetic field strength in the mid-plane as estimated with the revised equipartition formula by \citet{beck_05a};}
 \tablenotetext{e}{Rotation speed mostly obtained from the HyperLEDA extra-galactic data base;}
 \tablenotetext{f}{Radio haloes with cosmic-ray transport identified either as diffusion-dominated (\emph{Diff}) or advection-dominated (\emph{Adv});}
 \tablenotetext{g}{Best-fitting advection speed with $1\sigma$ uncertainties.}
 \tablecomments{It was assumed that the advection speed is constant in each of the galaxies for consistency (using $\tt velocity\_field = 0$, see Section~\ref{s:spinnaker}). For IC~10, NGC~891, NGC~3556, and NGC~4013 we also fitted optionally accelerating winds (see references).}
 \end{table*}

\subsubsection{Accelerating advection speed}

If the outflow has no lateral expansion, an accelerating wind can be a way to ensure energy equipartition in the halo. We notice that radio haloes have a box-shaped outline, where the radial extent of the halo hardly changes with height and is well correlated with the size of the star-forming disc \citep{dahlem_06a,heesen_18c,heald_21a}, which argues against a strong lateral expansion. Hence, dropping the assumption of a constant advection speed, we are able to ensure energy equipartition in the halo, for instance by using an exponential velocity distribution ($\tt velocity\_field =1$). This introduces one more free parameter, the velocity scale height $h_v$, so that the advection speed becomes $v(z)=v_0\exp(z/h_v)$. Galactic winds essentially accelerate linearly in the region where mass and energy is injected before the acceleration tailors off. This is the basic picture by the analytic wind model of \citet{chevalier_85a}. Including different driving agents such as radiation pressure and cosmic rays change this picture only slightly \citep{yu_20a}. \citet{heesen_18c} applied such a model successfully to the dwarf irregular galaxy IC~10. For exponential magnetic fields, energy equipartition requires $h_v\approx h_B/2$, so that the magnetic energy density is in agreement with the cosmic-ray energy density.

Another option is a wind velocity profile with a polynominal shape ($\tt velocity\_field=2$), where the advection velocity is parametrised as:
\begin{equation}
    v = v_0 \left [ 1 + \left (\frac{z}{h_v}\right )^\beta\right ].
    \label{eq:advection_profile}
\end{equation}{}
For $\beta=1$, the wind is linearly accelerating, whereas for $\beta=0.5$ the wind accelerates fast in the beginning and then the acceleration tailors off. The former is a good approximation for a cosmic ray-driven wind, where both simulations \citep{girichidis_18a} and semi-analytical 1D wind models \citep{everett_08a} predict linear velocity profiles. The latter is a closer approximation to stellar-driven wind models \citep{lamers_99a}.

\subsubsection{Advection in a wind}

Acceleration is not the only way to achieve equipartition in the halo, the second possibility is lateral expansion. Such a geometry can be a spherical outflow, as is the case with M~82, or a flux tube geometry which has been used to model cosmic ray-driven winds. We use the latter as this better represents the morphology of radio haloes. There is a choice of magnetic field parametrisation with either a pure vertical field geometry or a helical field with both azimuthal and vertical components. Faraday rotation measurements indicate that the magnetic field in the halo may be \emph{helical} \citep{heesen_11a,mora_19b,stein_20a}, so that there is an azimuthal component as well, hence we chose such a configuration. Nevertheless, we point out that there is a degeneracy between the assumed magnetic field geometry and the acceleration of the advection speed. Changing the magnetic field strength results in different energies of the CR$e^{-}$ we can probe (Equation~\ref{eq:cre_energy}), so that the spectral ageing is changed as well.

Hence, the third possibility is advection as a result of a simplified wind model using an iso-thermal wind solution (${\tt velocity\_field} =3$). This option will be motivated in more detail in Section~\ref{s:wind} \citep[see also][]{heald_21a}. Basically, this results in an approximately linear advection speed profile with approximate energy equipartition between the cosmic rays and the magnetic field. The simplified wind equation assumes a constant sound speed (iso-thermal wind model) and a flux tube geometry \citep{breitschwerdt_91a}. This allows us to describe a stellar feedback-driven wind with few free parameters; the parameters that need to be fitted are then advection speed at the critical point $v_0$, the flux tube scale height $z_0$ and the flux tube opening parameter $\beta$. This updated model is successful in matching the vertical distribution of non-thermal radio emission, and the vertical steepening of the associated spectral index, in
a consistent conceptual framework with few free parameters.

%
\begin{figure*}[tb]
\includegraphics[width=\textwidth]{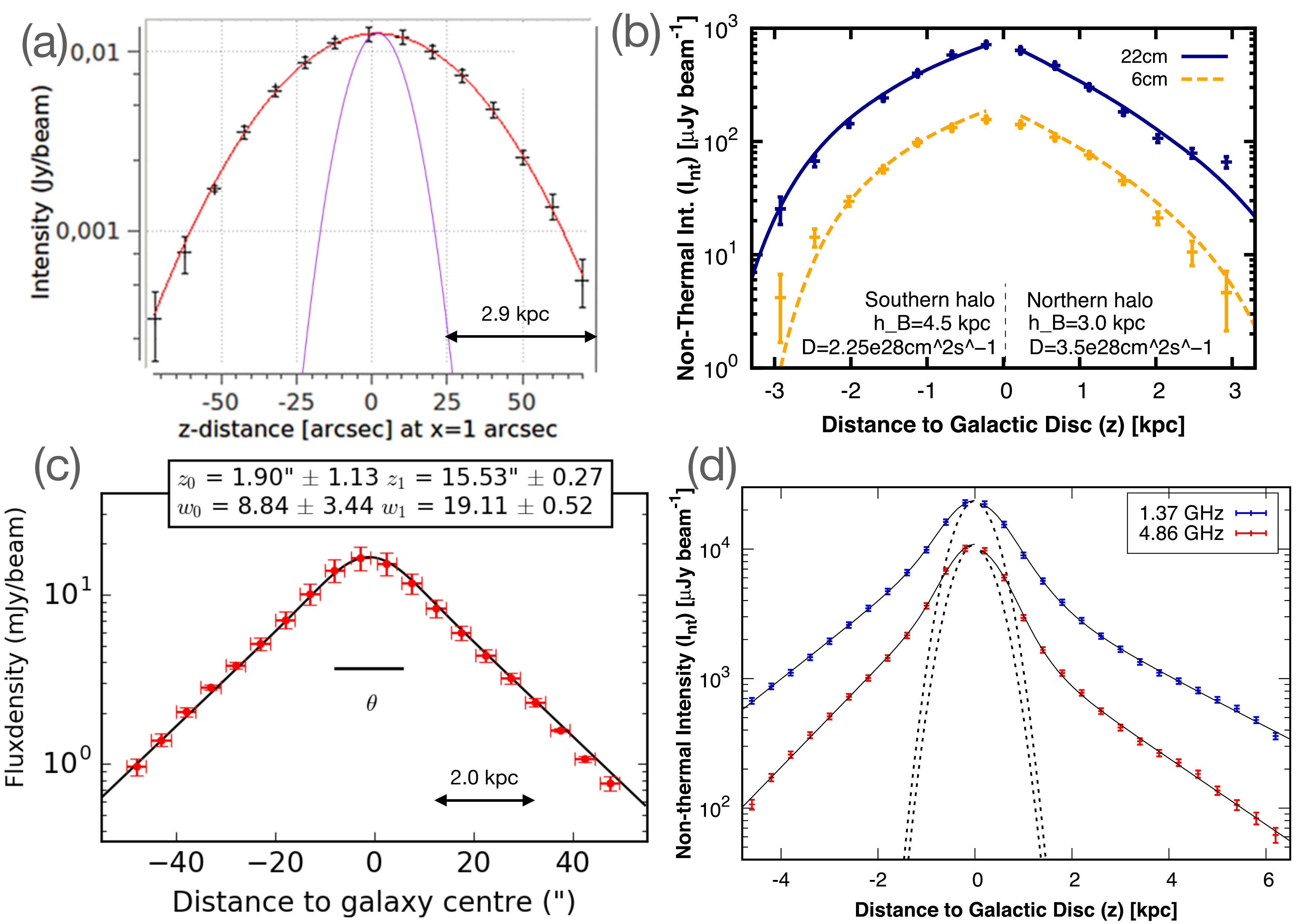}
\caption{Vertical intensity profiles in the haloes of four nearby edge-on spiral galaxies. The top row shows two examples for diffusion-dominated radio haloes with NGC~4565 at 140~MHz (a) and NGC~7462 at 1.4 and 4.7 GHz (b). The bottom row shows two examples which are advection dominated namely NGC 4217 at 140~MHz (c) and NGC~4631 at 1.4 and 4.9 GHz (d). Diffusion-dominated haloes have Gaussian intensity profiles (with preferentially only a 1-component profile), whereas advection-dominated radio haloes have exponential intensity profiles with preferentially 2-component profiles. Adopted from (a) \citet{heesen_19b}, (b) \citet{heesen_16a}, (c) \citet{stein_20a} and (d) \citet{heesen_18b}} 
\label{fig:diffusion}
\end{figure*}

\section{Radio haloes}
\label{s:radio_haloes}
Radio haloes offer us the possibility to apply the simple models of cosmic ray transport to the distribution of electrons in the halo. While some degeneracy remains between the magnetic field and the cosmic rays, the radio spectral index distribution and intensity distribution agrees to first degree with the models. This motivates to exploit the spatially resolved radio continuum emission to study cosmic-ray transport in more detail. Figure~\ref{fig:dumbbell} shows two prominent radio haloes as examples of what can be seen in the radio continuum. What is immediately clear is that the morphology of the radio haloes is not like a sphere, something that has been invoked to explain the radio sky background \citep{singal_15a}. With such an outside view we can also fairly easily check the size of the radio halo, as \citet{miskolczi_19a} could show the radio halo can extend to a size of up to 10~kpc as was also suggested by the modelling of the Milky Way halo \citep{orlando_13a}.

As this review focuses on what we have learned from the modelling with {\sc spinnaker}, we build on the sample by \citet{heesen_18b} who investigated 12 edge-on galaxies. Since then a few more galaxies were investigated in a similar way, so that we now have a sample of 16 galaxies that were analysed in a consistent way. In Table~\ref{tbl:sample}, these galaxies are listed. 

\subsection{Profile shape}
\label{ss:profile_shape}

Depending on the shape of the magnetic field distribution in the halo, the CR$e^{-}$ distribution is different for diffusion and advection, allowing us to distinguish between these two processes. Assuming an exponential magnetic field distribution is the first step since the radio continuum emission in the halo has this exponential distribution as well. Hence, the advection--diffusion approximation is used to show that diffusion leads to approximately \emph{Gaussian} intensity profiles and \emph{advection} leads to approximately \emph{exponential} intensity profiles \citep[see][and Section~\ref{s:spinnaker}]{heesen_16a}. 

\subsubsection{Gaussian profile shape}
\label{sss:gaussian_profile_shape}

Examples for Gaussian radio haloes with $I_{\nu}\propto \exp(-z^2/h_{\rm syn})$ are rare so far (see Fig.~\ref{fig:diffusion} (a) and (b)), with the only examples NGC~4013 \citep{stein_19a}, NGC~4565 \citep{heesen_19b} and NGC~7462 \citep{heesen_16a}. What these three galaxies have in common, however, are their low   star-formation rate surface densities with $\Sigma_{\rm SFR} < 2\times 10^{-3}~$~\usfrd. At these low values of \sfrd, simulations suggest that the formation of outflows is suppressed \citep{vasiliev_19a}. It is an exciting prospect that radio haloes can possibly establish whether such an outflow \sfrd-threshold really exists, and whether there are any other contributing factors such as a high mass-surface density. 

A possible \sfrd-threshold for the existence of gaseous haloes was posited already by \citet{rossa_03a}, who observed the extra-planar diffuse ionised gas (eDIG) in galaxies, which was later confirmed by X-ray observation of the hot ionised gas \citep{tuellmann_06a}. These observations suggested an \sfrd-threshold value similar to the one indicated by the diffusion--advection transition. The only other galaxy outside of this sample fitted with a single Gaussian component is NGC~4594 (M~104), an early type galaxy with a very low \sfrd\ \citep{krause_06a}, hence fitting the trend.

\subsubsection{Exponential profile shape}
\label{sss:exponential_profile_shape}

Most vertical intensity profiles are exponential so that $I_{\nu}\propto \exp(-z/h_{\rm syn})$ with either one or two components \citep{krause_18a}, showing that they are advection dominated. If there are two components, then we refer them to a thin and thick radio disc, respectively. This is in agreement with the finding that the scale heights are almost identical at both 1.5 and 6~GHz, suggesting an almost free escape of CR$e^{-}$ in a wind. Of the 16 galaxies considered in this review (see Table~\ref{tbl:sample}), 13 have exponential radio continuum profiles (Fig.~\ref{fig:diffusion}(c) and (d)). There are other galaxies outside of our sample, which have been fitted with exponential profiles, such as NGC~3034 \citep[M~82;][]{adebahr_13a}.

\subsubsection{Multi-component radio disc}
\label{sss:multi_component_radio_disc}

It is an open question whether galaxies always have both thin and thick radio discs, as is predominantly found by observations thus far. Generally speaking, our observations thus far indicate that galaxies have either a 2-component exponential vertical distribution, consisting of both and thick radio discs, or a 1-component Gaussian disc, consisting of a thick radio disc only. Of course, this will be resolution dependent since most thin radio discs have only a scale height of a few hundred parsec \citep{heesen_18b}, so that the angular resolution has to be sufficiently high in order to resolve them. In the sample discussed here, only 2 out of the 16 galaxies do not have a multi-component radio disc, NGC~4565 and NGC~7462, which both possess only a thick radio disc. It is notable that these two galaxies are diffusion-dominated. The only other galaxy outside of this sample that has a Gaussian vertical intensity profile is NGC~4594 (M~104), which is also fitted by a single Gaussian component \citep{krause_06a}. We can speculate that diffusion results in only a thick radio disc, whereas in the case of advection both the thin and thick discs form. Since diffusion dominates near the disc, the thin radio disc will be diffusion dominated and advection takes over as the dominating transport mode, where the profile flattens and the thick radio disc begins (Section~\ref{sss:cre_transport_length}). 

Such a transition in the cosmic-ray distribution is also seen in cosmological simulations with {\sc fire-2}, where the transition is at 10~kpc height \citep[which is expected as][use much larger diffusion coefficients]{hopkins_20a}. \citet{girichidis_18a} find a flattening of the profile at $0.5$~kpc height with a more typical diffusion coefficient of $10^{28}$~\udif. As equation~\eqref{eq:diff_adv} predicts, for typical advection speeds of a few 100~\uvel\ and diffusion coefficients of $10^{28}~\rm cm^2\,s^{-1}$, we expect the transition to happen at around 1~kpc or less. Thus we raise the possibility that a galaxy with a wind has a two-component radio disc, whereas no-wind galaxies have only a one-component radio disc with a thick disc. NGC~4013 is the only galaxy that has a two-component Gaussian radio disc; this galaxy is a hybrid case where diffusion and advection both contribute because the advection speed is sufficiently slow \citep{stein_19b}.

\subsection{Scale heights}
\label{s:scale_heights}

\subsubsection{Global measurements}
\label{ss:scale_heights}

%
\begin{figure}[tb]
\includegraphics[width=\columnwidth]{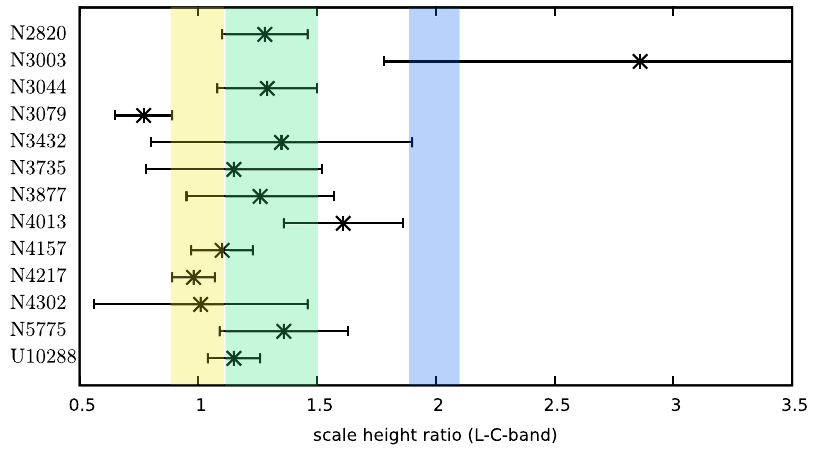}
\caption{The exponential radio continuum scale height in the CHANG-ES sample at $1.5$ and 6~GHz with the corresponding ratio. Shaded areas show the expectation for non-calorimetric advection (free escape; yellow), calorimetric diffusion (green), and calorimetric advection (blue). From \citet{krause_18a}} 
\label{fig:scaleheight}
\end{figure}

%
\begin{figure}[tb]
\includegraphics[width=\columnwidth]{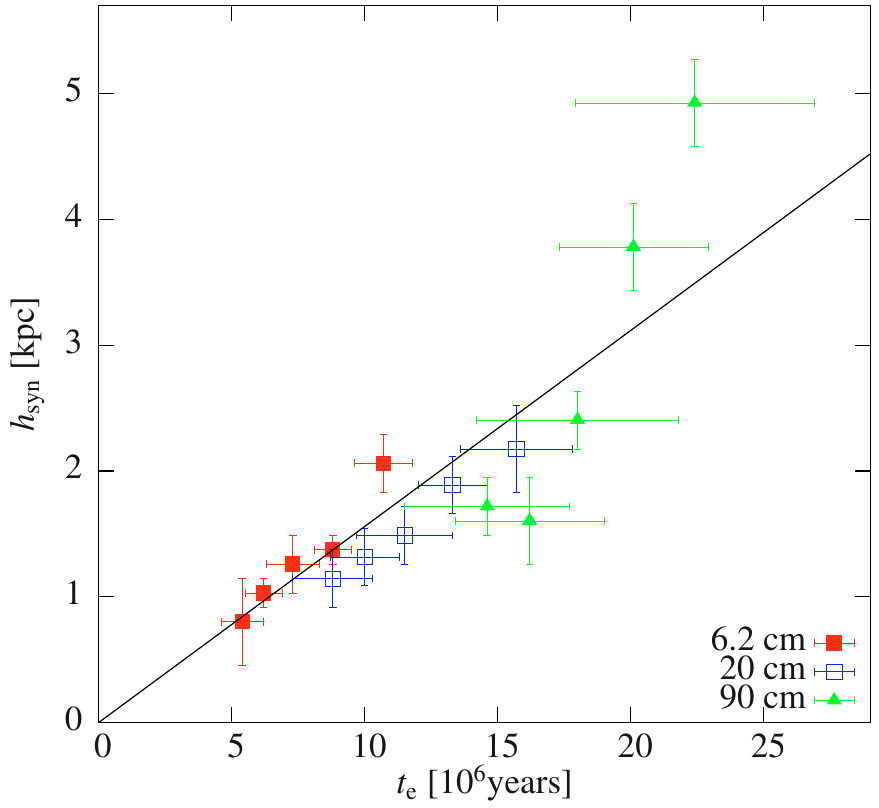}
\caption{The non-thermal exponential radio continuum scale height in NGC~253 as function of the CR$e^{-}$ synchrotron lifetime. The line shows the best-fitting advection solution. From \citet{heesen_09a}} 
\label{fig:hadvection}
\end{figure}

In Fig.~\ref{fig:scaleheight}, we present the scale height ratio between $1.5$ and 6~GHz in the CHANG-ES sample \citep{krause_18a}. For advection, we would expect the ratio to be 2 (equation~\ref{eq:advection_scale_height}), for diffusion to be around $1.3$ (depending on $\mu$; equation~\ref{eq:diffusion_scale_height}), and for a free escape we would expect the ratio be 1. As can be seen, the ratio is in agreement with either diffusion with a significant energy loss or free escape. What we can rule out, however, is advective transport in a calorimetric halo, although diffusive transport in a calorimetric halo would be still possible. However, there are two reasons that argue against this latter option: first, the exponential profiles are in agreement with advection;  second, the galaxies have integrated radio spectral indices that are not steep enough in order classify them as CR$e^{-}$ calorimeters. Hence, the scale height analysis points to advective transport in winds \citep{krause_18a}.

\subsubsection{Spatially resolved measurements}

The second method using scale heights to measure CR$e^{-}$ transport, is to use spatially resolved measurements. For a given galaxy, the mode of CR$e^{-}$ transport should not change much across the size of galaxy, for instance in a galaxy-wide outflow advection dominates. In this case, the \emph{local} scale height will be a function of the local CR$e^{-}$ lifetime, which depends on the local magnetic field strength. The motivation for this approach was the observation of radio haloes that have a `dumbbell' shape, meaning smaller radio scale heights in the centre of the galaxy and increasing scale heights in their outskirts. Examples for this type of haloes are NGC~253 (Fig.~\ref{fig:dumbbell}(a)), NGC~891 (Fig.~\ref{fig:dumbbell}(b)) and NGC~4217 \citep{stein_20a}. 

\begin{figure}[tb]
\includegraphics[width=\columnwidth]{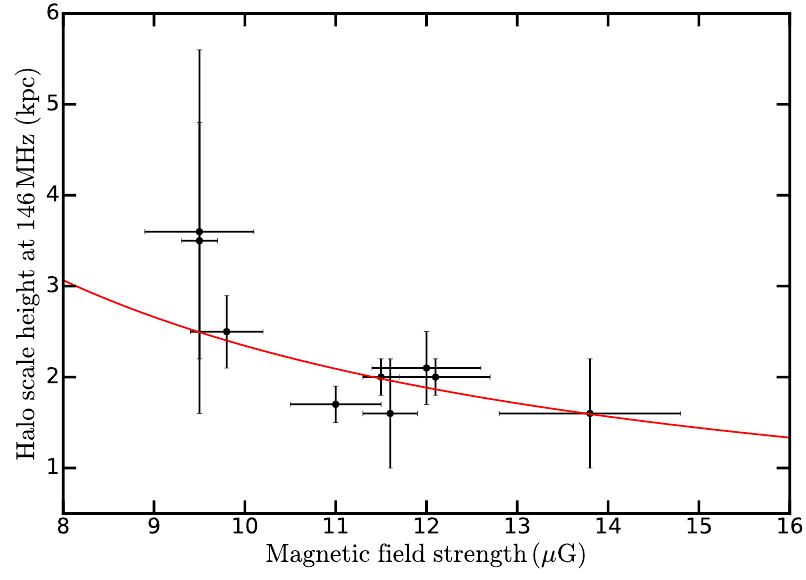}
\caption{The radio continuum exponential scale height at 146~MHz in NGC~891 as function of the magnetic field strength in the mid-plane. The red line shows the best-fitting exponential function. From \citet{mulcahy_18a}} 
\label{fig:n891_h}
\end{figure}

The CR$e^{-}$ scale height can be compared with expected relations for advection (equation~\ref{eq:advection_scale_height}) and diffusion (equation~\ref{eq:diffusion_scale_height}). The first measurement of cosmic-ray advection with this method of comparing the CR$e^{-}$ distribution with the magnetic field strength was presented by \citet{heesen_09a}, who found that the radio continuum scale height scales linearly with the CR$e^{-}$ lifetime as presented in Fig.~\ref{fig:hadvection}. Consequently, they calculated the cosmic-ray advection speed to be $v=300\pm 30$~\uvel. The alternative is to study directly the dependence of the CR$e^{-}$ scale height on the magnetic field field strength. This has been done by \citet{mulcahy_18a} for NGC~891, who found a dependence of $h_{\rm e}\propto B^{-1.2\pm 0.6}$, in agreement with either diffusion or advection (see Fig.~\ref{fig:n891_h}).

\subsection{Size--scale height relation}
\label{ss:size_scale_height_relation}
\citet{krause_18a} studied the scale heights in CHANG-ES galaxies and found that the scale height scales linearly with the size of the galaxy. In order to exclude the size of the galaxy, they defined a normalised scale height. This normalised scale height fulfils a scale height--mass surface density relation, where the normalised scale height decreases with increasing mass-surface density. Both relations point to a relation of the radio halo with stellar feedback. Interestingly, both the intensity and magnetic field scale height do not depend on either the SFR, \sfrd, or rotation speed \citep{heesen_18b}. This might point to a geometric model with an expanding outflow as well, as do the results of \citet{krause_18a}.

%

%
\begin{figure}[tb]
\includegraphics[width=\columnwidth]{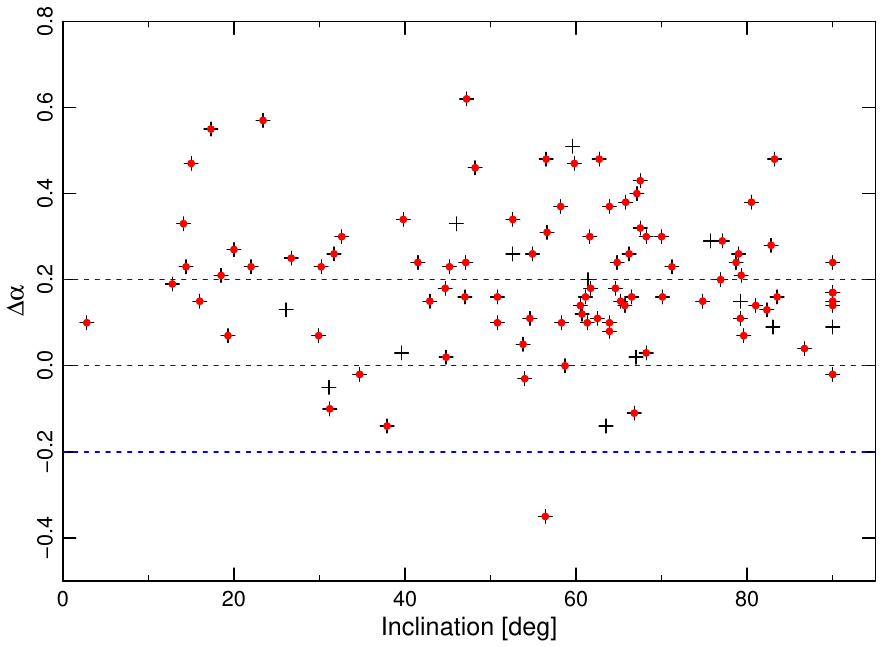}
\caption{Spectral curvature as function of the inclination angle. The differential spectral index is defined as $\Delta\alpha = \alpha_{\rm low}-\alpha_{\rm high}$, where $\alpha_{\rm low}$ is the radio spectral index between 150 and 1500~MHz and $\alpha_{\rm high}$ is the radio spectral index between 1500 and 5000~MHz. From \citet{chyzy_18a} } 
\label{fig:alpha}
\end{figure}

%
\begin{figure}[!tbh]
\includegraphics[width=\columnwidth]{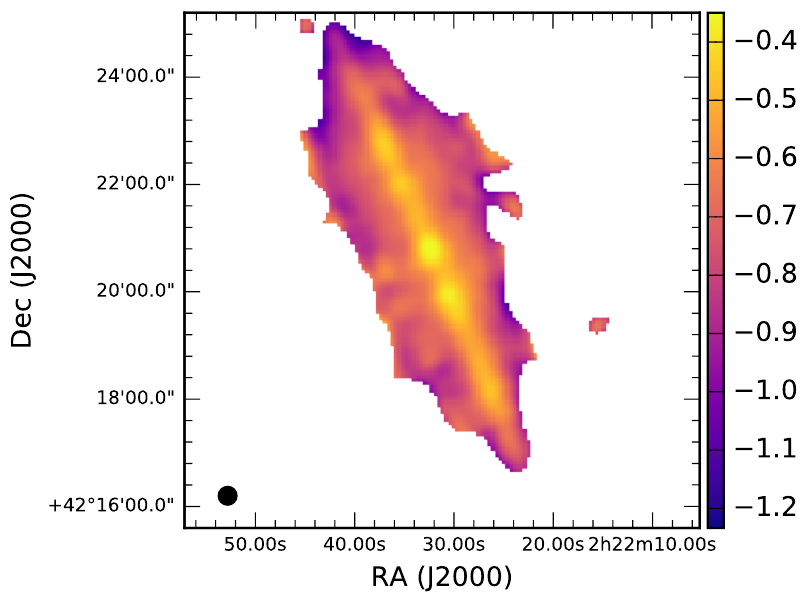}
\caption{The radio spectral index  in NGC~891 between 146 and 1500~MHz. The spectral index is flat in the disc (although clearly non-thermal) and steepens in the halo. From \citet{mulcahy_18a} } 
\label{fig:n891_alpha}
\end{figure}

\section{Radio continuum spectrum}
\label{s:radio_continuum_spectrum}

\subsection{Global spectrum}
\label{ss:global_spectrum}

Observations show the integrated (global) radio continuum spectrum of galaxies to be in agreement with a power-law with a non-thermal radio spectral index of $-0.9$ at frequencies between 1 and 10~GHz \citep{tabatabaei_17a}. However, at low frequencies ($<1~\rm GHz$) the radio continuum spectrum deviates from a power-law and the spectrum flattens significantly \citep{marvil_15a}. The most comprehensive study to date is that of \citet{chyzy_18a}, who studied $\sim$100 galaxies with LOFAR and archival data between 50~MHz and 5~GHz. They found that the spectral index flattens by $\Delta\alpha = 0.2$ from a spectral index of $\alpha=-0.77$ above $1.5$~GHz to $\alpha=-0.57$ below $1.5$~GHz. Hence the low-frequency spectral index is close to the injection spectral index, which means that the CR$e^{-}$ may be able to escape the galaxy freely. This view is supported by the observation that the steepening of the spectrum is independent of the inclination angle (see Fig.~\ref{fig:alpha}). Prior, it was posed that internal effects such as due to \emph{free--free absorption} at low frequencies, the spectrum is artificially flattened \citep{israel_90a}. This interpretation seems to be now at least unlikely.


\subsection{Spatially resolved measurements}

In edge-on galaxies, the radio spectral index can be also measured locally. Since the radio spectral index is fairly flat in the disc ($\alpha\approx -0.6$), the star-forming galactic mid-plane, this suggests that the CR$e^{-}$ are freshly injected. In the halo, the radio spectral index steepens to values of $\alpha\approx -1$ or even steeper (see Fig.~\ref{fig:n891_alpha}).

%
\begin{figure}[tb]
\includegraphics[width=\columnwidth]{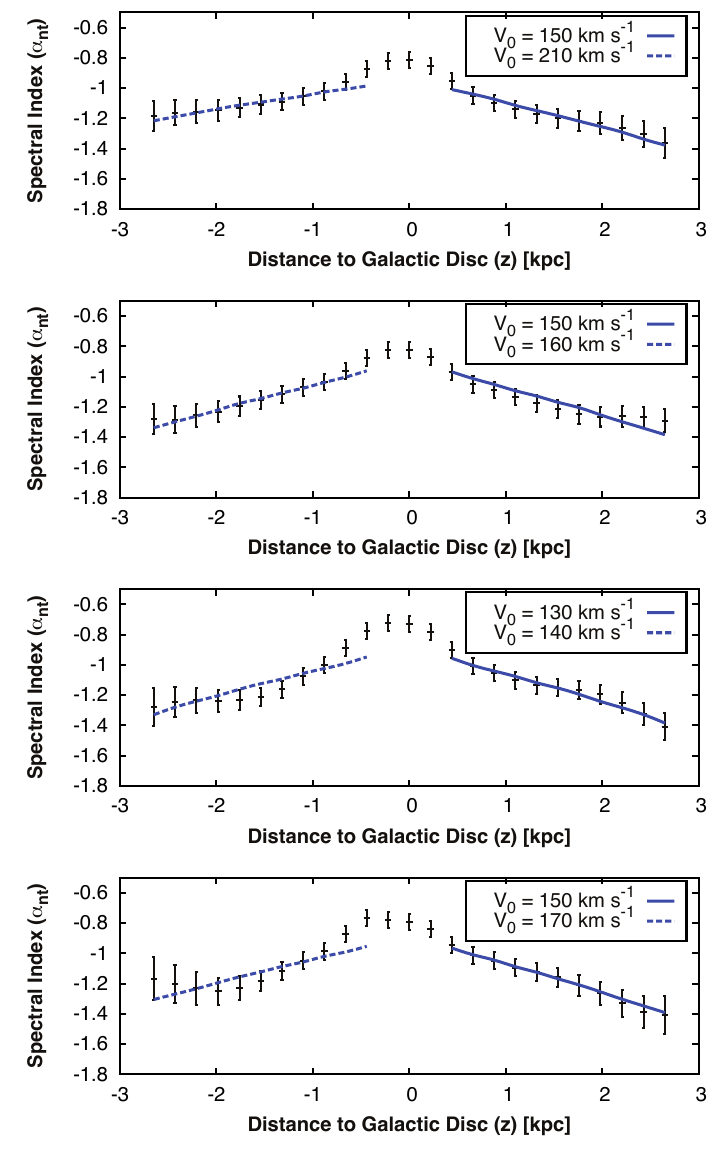}
\caption{Vertical profiles of the non-thermal radio spectral index between 146 and 1500~MHz in NGC~891. Lines show best-fitting advection models, which well describe the linear decrease of the spectral index in the halo at $|z|\gtrsim 0.5$~kpc. From  \citet{schmidt_19a}} 
\label{fig:n891_alpha_profile}
\end{figure}
%
\begin{figure}[!htb]
\includegraphics[width=\columnwidth]{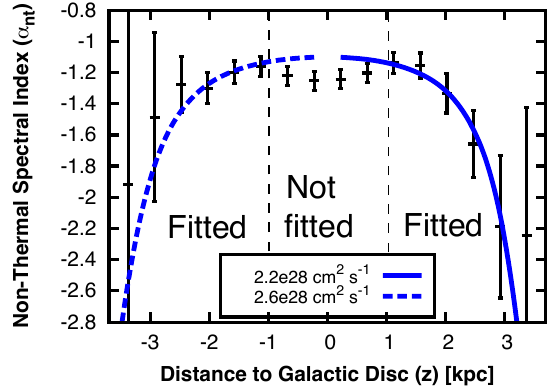}
\caption{Vertical radio spectral index profile between $1.4$ and $4.7$~GHz in NGC~7462 as an example for diffusion-dominated transport. The profile has a `parabolic' shape, lines show the best-fitting diffusion model, which was fitted in the halo only at $|z|>1~\rm kpc$. From \citet{heesen_16a} } 
\label{fig:n7462_diffusion}
\end{figure}

The next step is to use the advection--diffusion approximation to calculate vertical radio spectral index profiles. In Fig.~\ref{fig:n891_alpha_profile}, we present vertical spectral index profiles in NGC~891, which are approximately linear. The spectral index profiles show a flat spectral index in the disc, rapidly steepening in the halo, so that one finds a two-component spectral index profile. In contrast, in Fig.~\ref{fig:n7462_diffusion} we present the vertical radio spectral index profile for NGC~7462, a diffusion-dominated galaxy. In this case, the spectral index is already quite steep in the disc with values of $\alpha\approx -1.2$, as would be expected for a calorimetric galaxy, with no CR$e^{-}$ escape. Remarkably, the radio spectral index does not steepen out to distances of $z\approx 2$~kpc, quite differently to advective galaxies. The best other example for this kind of radio spectral index profiles is NGC~4565 \citep{schmidt_19a}, which has also remarkably steep spectral indices in the disc. Hence, we indeed find vertical spectral index profiles in approximate agreement with our idealised versions of the pure diffusion and advection models (Section~\ref{s:spinnaker}).

This then motivates the application of the {\sc spinnaker} models to the edge-on galaxies to decide whether they are diffusion- or advection-dominated and to measure diffusion coefficients and advection speeds (see Table~\ref{tbl:sample}). We will return to these results in Section~\ref{s:results}.

%
\begin{figure*}[tb]
\includegraphics[width=\textwidth]{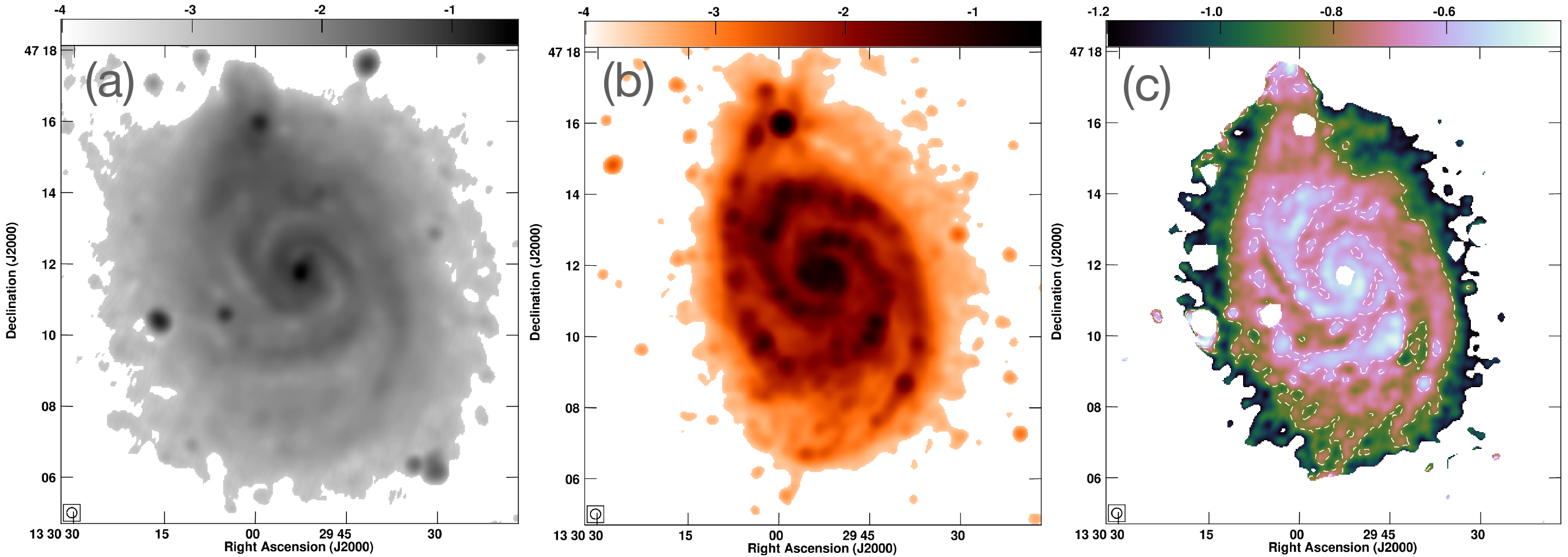}
\caption{Example for the influence of CR$e^{-}$ transport in the face-on galaxy NGC~5194. The 145-MHz radio continuum map (a) is the smoothed version of the \sfrd-map (b). Both maps are showing the star-formation rate surface density as $\log_{10}(\Sigma_{\rm SFR}/{\rm M_\odot\,yr^{-1}\,kpc^{-2}})$. (c) shows the radio spectral index between 145 and 1365~MHz. From \citet{heesen_19a}} 
\label{fig:smoothing}
\end{figure*}

\section{Face-on galaxies}
\label{s:face_on_galaxies}
\subsection{Smoothing experiments}
\label{ss:smoothing_experiment}
A different approach of measuring the cosmic-ray transport length is to study face-on galaxies. The radio continuum emission is smoothed with respect to the star-formation rate surface density (\sfrd), which can be explained with CR$e^{-}$ transport (Fig.~\ref{fig:smoothing}). This idea was first exploited by \citet{murphy_08a} who compared the $1.4$-GHz emission from the WSRT--SINGS sample by \citet{braun_07a} with 70-$\mu$m far-infrared emission from \emph{Spitzer}. They convolved the far-infrared map with both exponential and Gaussian kernels and minimised the difference between the convolved map and the radio continuum map. The half-length of the smoothing kernel is then the cosmic-ray transport length, which lies between $0.4$ and $2.3$~kpc. They found the length to be a function of the \sfrd, which can be explained by increased synchrotron and IC losses and thus shorter CR$e^{-}$ lifetimes.

The same approach was used in \citet{vollmer_20a}, who investigated both $1.4$- and 5-GHz radio continuum maps. They found that the cosmic-ray transport length is a function of frequency with $1.8\pm 0.5$~kpc at $1.4$~GHz and $0.9\pm 0.3$~kpc at 5 GHz. They also tested both exponential and Gaussian kernels and found that the goodness of the fit cannot be used to distinguish between advection (streaming) and diffusion. However, they found that in several galaxies the $1.4$/5~GHz-ratio of the transport length is larger than $1.5$, an indication for streaming (equation~\ref{eq:advection_scale_height}). This interpretation not dependent on the question of electron calorimetry since escape would lead to an even smaller frequency dependency. Ideally, one would like to measure the shape of the CR$e^{-}$ transport kernel in order to make a distinction between different models, but so far exponential and Gaussian kernels cannot be distinguished by their fitting quality alone \citep{murphy_08a,vollmer_20a}. This is easier to do in edge-on galaxies since we can measure the shape directly assuming that the CR$e^{-}$ are injected only in the thin star-forming disc. 

\subsection{Radio--SFR relation}
\label{ss:radio_sfr}
A variation of the smoothing experiment (Section~\ref{ss:smoothing_experiment}) is to study the spatially resolved radio--SFR relation, where we plot the radio continuum emission as a function of the \sfrd-values \citep{berkhuijsen_13a}. The radio continuum--star-formation rate (radio--SFR) relation is approximately linear for global measurements \citep{heesen_14a}, but the spatially resolved radio--SFR relation is sub-linear with slopes of $0.6$ when measured at 1-kpc spatial resolution. The \sfrd-map can then be convolved with a Gaussian kernel in order to linearise the radio--SFR relation \citep{berkhuijsen_13a,heesen_14a,heesen_19a}. \citet{heesen_19a} found that the half-width of the Gaussian kernel, the cosmic-ray transport length, is a function of frequency. Depending on the frequency-dependence, the transport is dominated by either by cosmic-ray diffusion or streaming.

\begin{figure}[tb]
\includegraphics[width=\columnwidth]{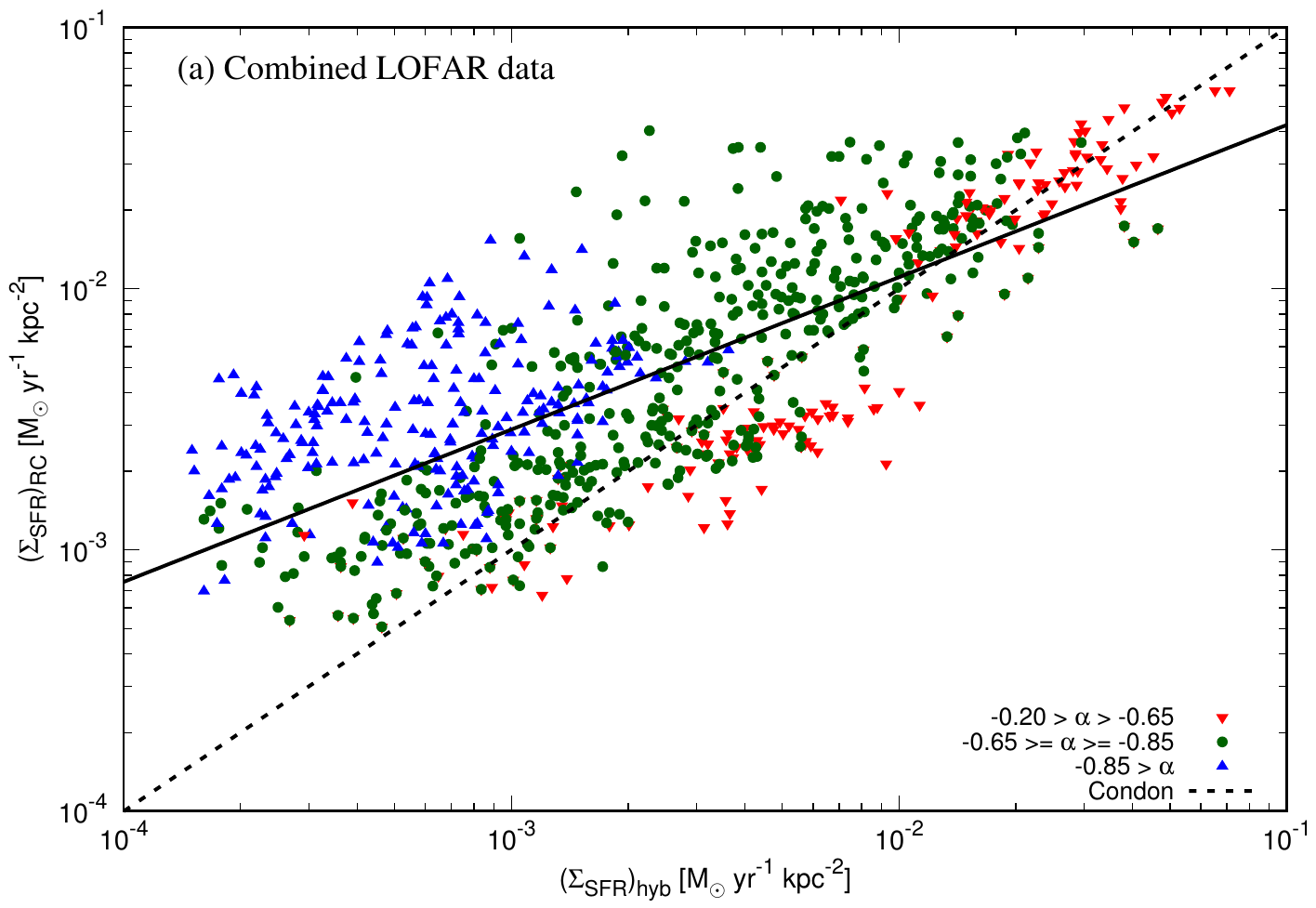}
\caption{Spatially resolved radio--SFR relation in four late-type galaxies at 145~MHz. The radio-derived SFR surface density is here shown as function of the mid-infrared and far-ultraviolet hybrid SFR surface density. Data points are coloured according to their radio spectral index, with red data points indicating young CR$e^{-}$, green points CR$e^{-}$ of intermediate age, and blue points old CR$e^{-}$. The solid line shows the best-fitting relation with a sub-linear slope (when compared with the dashed 1:1 relation) that can be attributed to cosmic-ray transport. From \citet{heesen_19a}} 
\label{fig:m51_sfr}
\end{figure}

The key finding of the spatially resolved radio--SFR relation is that the deviation from the theoretical expectation such as the Condon relation \citep{condon_92a} and its more recent derivatives \citep{murphy_11a} is dependent on the radio spectral index. For a fairly flat radio spectral index of $\alpha\approx -0.6$, the deviation is small and so the relation is almost linear (red data points in Fig.~\ref{fig:m51_sfr}). This fits our expectation that on a kpc-scale the  radio--SFR relation is linear as long as the CR$e^{-}$ are young and cosmic-ray transport plays no role. \citet{dumas_11a} and \citet{basu_15a} found linear radio--SFR relations in the spiral arms of galaxies, where the spectral index is flat as well. Contrary, if the radio spectral is steep $\alpha<-0.85$, the radio continuum emission lies above the radio--SFR relation. This finding is important because it shows that spectral ageing is important shaping the relation. In areas of low star-formation rates, old CR$e^{-}$ have diffused into these areas and thus boost the radio continuum emission above the level of what would be expected for the local \sfrd.

%

%
\begin{figure}[tb]
\includegraphics[width=\columnwidth]{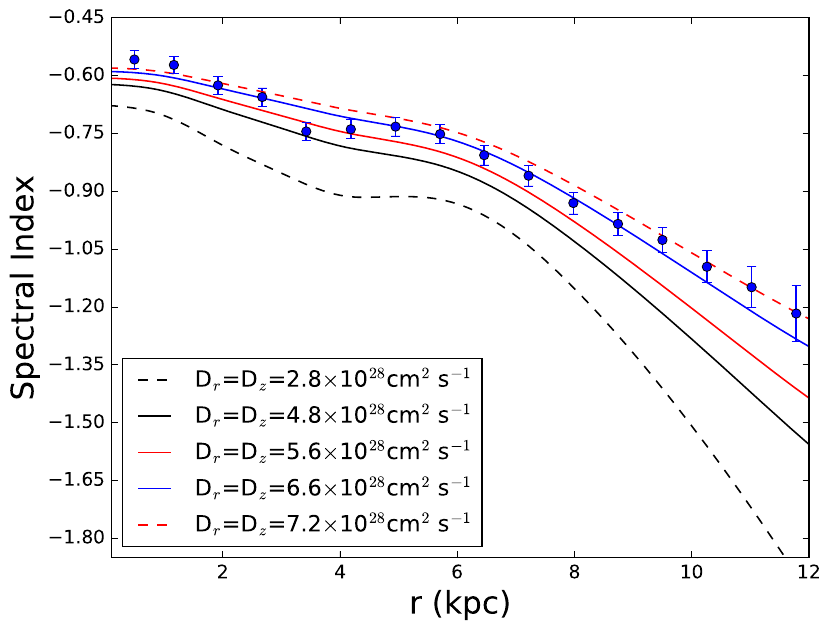}
\caption{The radial radio spectral index profile in NGC~5194 (M~51) between 140 and 1500~MHz. Lines show various 1D diffusion models. From \citet{mulcahy_16a}} 
\label{fig:m51_alpha}
\end{figure}

\subsection{Diffusion modelling}

\citet{mulcahy_16a} solved the diffusion--loss equation for radial transport of CR$e^{-}$ in the 1D case. The radial intensity profiles and spectral index profiles are fitted with the diffusion coefficient and its energy dependency. They found that a diffusion coefficient of $6\times 10^{28}~\rm cm^2\,s^{-1}$ as their best-fitting solution. Their energy-dependence is consistent with zero, meaning that the diffusion coefficient is constant. Fig.~\ref{fig:m51_alpha} shows the radial spectral index profile that they modelled. A key finding is that the spectral index is too steep without escape of CR$e^{-}$. Hence, they included the diffusive escape time as:
\begin{equation}
    t_{\rm esc} = \frac{h^2}{D}.
\end{equation}
They used H\,{\sc i} scale heights to measure $h$, which are between 3 and 9 kpc, so that the escape time is between 11 and 88~Myr. The spectral index profile in Fig.~\ref{fig:m51_alpha} shows that the best-fitting solution. It can be seen that the model shows a smaller radial  variation than the observed data, in particular the minimum at $r=4~\rm kpc$, so that a better fit might be obtained with a smaller diffusion coefficient and escape in a wind.

\section{Results}
\label{s:results}

In this section, we summarize the results that have so far been obtained for the cosmic-ray transport in external galaxies using radio continuum observations.

\subsection{Diffusion coefficients}
\label{ss:diffusion_coefficients}

The measured diffusion coefficients are between values of $10^{27}$ and $10^{29}$~\udif, with most values at around $10^{28}$~\udif\ \citep{murphy_08a,murphy_11a,berkhuijsen_13a,heesen_18b,heesen_19b,vollmer_20a}. This is expected since we are tracing a few kpc-scales and the CR$e^{-}$ lifetime is a few 10~Myr, resulting in this number using equation~\eqref{eq:diffusion_coefficient}. The lowest diffusion coefficients are found in dwarf galaxies \citep{murphy_11a,heesen_18c} with the highest ones in radio haloes \citep{heesen_09a,heesen_18b}. The diffusion coefficients depend weakly on the far-infrared (SFR) surface density as \citet{murphy_08a} have shown. This is expected as long as the CR$e^{-}$ lifetime is dependent on $\Sigma_{\rm SFR}$, because $t_{\rm syn}\propto B^{-3/2}$ and $B\propto \Sigma_{\rm SFR}^{1/3}$, we expect $t_{\rm syn}\propto \Sigma_{\rm SFR}^{-1/2}$. Thus, the diffusion length should be $L\propto \Sigma_{\rm SFR}^{-1/4}$ for a non-energy dependent diffusion coefficient. This is in approximate agreement with the results of \citet{murphy_08a} although the scatter is quite significant. \citet{tabatabaei_13a} repeated this experiment and found no dependence on the SFR surface density, although their sample was fairly small.

\subsubsection{Energy dependence}
\label{sss:energy_dependence}
The energy dependence of the diffusion coefficient has been explored as well. There are cases when no energy dependence is needed to fit the data, such as is the case if the diffusion length scales as $L\propto t_{\rm syn}^{-1/2}$ or, expressed as frequency, $L\propto \nu^{-1/4}$ (Section~\ref{ss:expected_relations}). Frequently, the frequency dependence of the diffusion length $L(\nu)$ is flatter such as $L\propto \nu^{-1/8}$, but this can be also a result of electron non-calorimetry \citep{heesen_19a}. The edge-on galaxy so far analysed in most detail with a pure diffusion halo, NGC~4565, is indeed better consistent with a energy independent diffusion coefficient or only a weakly dependent diffusion coefficient \citep{heesen_19b, schmidt_19a}. Essentially, an energy-dependent diffusion coefficient would lead to an even more pronounced curvature of the radio spectral index profile than what is observed. \citet{heesen_16a} tested the energy-dependence in NGC~7462, but did not find a strong indication for it. 

A different approach is to fit the Gaussian convolution kernel in face-on galaxies with a cosmic-ray diffusion model. \citet{heesen_19a} did this and found the energy dependence to vary widely with $\mu=0$--$0.6$. There are indications, however, in particular from the radio spectral index, that high values of $\mu$ are the result of CR$e^{-}$ escape (electron non-calorimetry) rather than an intrinsic feature of the CR$e^{-}$ transport. In summary, diffusion coefficients in the GeV-range seem to be not energy dependent. In those cases where we see a dependence, the indication is either weak (in edge-on galaxies) or can be largely explained by flat radio spectral indices hinting at CR$e^{-}$ escape (in face-on galaxies). The observation that for a few GeV the diffusion coefficient is not energy-dependent is in agreement with the Boron-to-Carbon (secondary to primary) cosmic-ray ratio in the Milky Way \citep{becker_tjus_20a}. 


\subsection{Cosmic-ray streaming}
\label{ss:cosmic_ray_streaming}

The indications for cosmic-ray streaming come mostly from scaling of the CR$e^{-}$ transport length with frequency, which in case of streaming resembles advection rather than diffusion (Section~\ref{ss:expected_relations}). \citet{vollmer_20a} found two galaxies where the CR$e^{-}$ transport length scales more with the frequency than can be explained by pure diffusion even when the diffusion coefficient is assumed to be energy independent. Similarly, \citet{beck_15b} found in IC~342 the CR$e^{-}$ propagation length to scale with $L\propto \nu^{-0.5}$. This can be explained by cosmic-ray streaming, where the CR$e^{-}$  are transported with a constant speed, for instance the Alfv\'en speed. We may consider the influence of an advection-dominated radio halo, which would result in a similar behaviour. What argues against such a halo is that an advective halo will limit the confinement of cosmic rays, which would again limit the effective CR$e^{-}$ lifetime and thus reduce the frequency dependence. Taken together, the results by \citet{beck_15b} and \citet{vollmer_20a} seem to be strongly indicative of cosmic-ray streaming. Another hint comes from \citet{tabatabaei_13a} who found that the CR$e^{-}$ transport length in NGC~6946 is larger than what one would expect from the ratio of ordered and turbulent magnetic field strength. 

In edge-on galaxies, cosmic rays can stream from the disc into the halo along  vertical magnetic field lines. Obviously, in diffusion-dominated galaxies streaming must be suppressed, so that we can assume that galaxies without outflows do not have the right type of magnetic field structure, presumably lacking vertical magnetic field lines. Indeed, the two pure diffusion haloes in our sample, NGC~4565 and NGC~7462, have no  dominant vertical magnetic field lines \citep{heesen_16a,wiegert_15a}. The hybrid diffusion--advection galaxy NGC~4013 has at least a significant vertical magnetic field component \citep{stein_19a}. In galaxies with winds, advection and streaming may be observed together although a separation of them is difficult. In the edge-on galaxy NGC~5775, the vertical radio spectral index gradient is much reduced at the position of vertical magnetic field lines \citep{duric_98a,heald_21a}. This could also be the result of CR$e^{-}$ streaming; the effective CR$e^{-}$ bulk speed is then the superposition of Alfv\'en and wind speed.

%
\begin{table}[tb]
\small
\caption{Advection speed scaling relations} 
 \label{tbl:scale}
\begin{tabular}{lc}
 \tableline  
 Advection speed in \uvel & $\rho_{\rm s}$ \\
  \tableline  
  $v =10^{2.13\pm 0.05} (SFR/{\rm M_\odot\,yr^{-1}})^{0.41\pm 0.06}$ & $0.75$ \vspace{0.1cm} \\
  $v = 10^{3.23\pm 0.25}(\Sigma_{\rm SFR}/{\rm M_\odot\,yr^{-1}\,kpc^{-2}})^{0.41\pm 0.13}$ & $0.70$\tablenotemark{a} \vspace{0.1 cm} \\
  $v =10^{2.085\pm 0.056}(v_{\rm rot}/{100~\rm km\,s^{-1}})^{1.40\pm 0.20}$ & $0.54$ \\ 
  \tableline 
 \end{tabular}
%
 \tablenotetext{a}{IC~10 excluded from fit}
 \end{table}

%
\begin{figure*}[!thb]
\centering
\includegraphics[width=\textwidth]{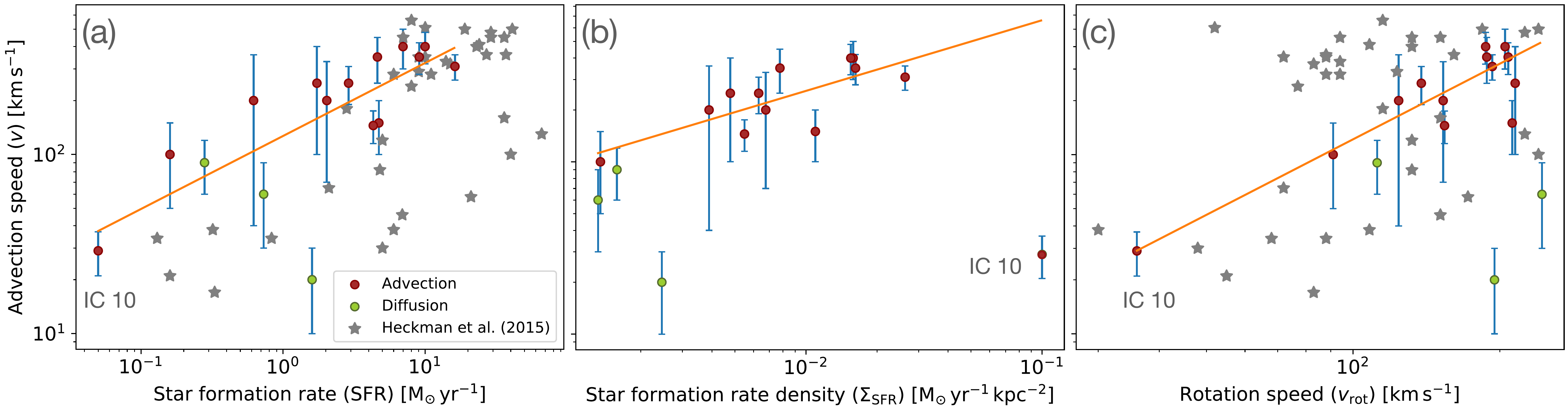}
\caption{Re-evaluated scaling relations of the CR$e^{-}$ advection speed in edge-on galaxies. (a) shows the advection speed as function of SFR, (b) as function of SFR surface density and (c) as function of the rotation speed. Best-fitting advection scaling relations (Table~\ref{tbl:scale}) are shown as solid lines. Values for the filled circular data points are from Table~\ref{tbl:sample}; grey stars show UV absorption line measurements for a different sample of galaxies which were taken from \citet{heckman_15a}} 
\label{fig:scaling}
\end{figure*}

\subsection{Anisotropic diffusion}
\label{ss:anisotropic_diffusion}

The question whether diffusion happens isotropic or anisotropic is of importance for the modelling of galaxy evolution. \citet{vollmer_20a} used elliptical smoothing kernels aligned with the magnetic field as measured from linear polarisation and found slight indication that the CR$e^{-}$ are preferentially transported along magnetic field lines. An indirect way to study the influence of the magnetic field may be using the radio spectral index as a proxy for CR$e^{-}$ confinement times. In face-on galaxies, we find steep radio spectral indices in  inter-arm regions with strong ordered magnetic fields. Such areas may be the places where the CR$e^{-}$ are stored by disc-parallel magnetic fields, before they can escape into the halo. Prominent examples are NGC~5055 \citep{heesen_19a}, NGC~5194 \citep[M~51;][]{mulcahy_14a} and NGC~6946 \citep{tabatabaei_13a}. Corroborating the influence of the magnetic field, galaxies lacking a large-scale spiral magnetic field, such as the dwarf irregular galaxy IC~10, show a flat spectral index throughout the disc \citep{heesen_18c}. 

In NGC~253, \citet{heesen_11a} found that the CR$e^{-}$ diffusion across a magnetic filament perpendicular to the field direction is quite fast, with a diffusion coefficient of $D_\perp = 1.5\times 10^{28}~\rm cm^2\,s^{-1}$. This is a fairly high diffusion coefficient for pure perpendicular diffusion, which can be explained by a small amount of turbulence in the magnetic field. One can also take the radio haloes as a proxy for anisotropic diffusion. In this case diffusion coefficient tend to be quite high of the order $10^{29}~\rm cm^2\,s^{-1}$ \citep{dahlem_95a, heesen_09a}. \citet{buffie_13a} provided a theoretical explanation for the ratio of the perpendicular to parallel diffusion coefficient, which involves the turbulent component of the magnetic field which can be described by the so-called correlation length (similar to the field line bend-over length).

\subsection{Advection speed scaling relations}
\label{ss:scaling_relations}
The advection speed scaling relations with SFR, \sfrd, and the rotation speed $v_{\rm rot}$ were already investigated by \citet{heesen_18b}. For this review, we have re-evaluated their sample which we extended to 16 galaxies (Table~\ref{tbl:sample}). In our sample, three galaxies are diffusion-dominated, which we exclude in the fitting process but present them in the plots for comparison. In Table~\ref{tbl:scale} an overview of the scaling relations discussed can be found.

The advection speed as function of the SFR is presented in Fig.~\ref{fig:scaling}(a), where the advection speed scales with the SFR as $v\propto SFR^{0.4}$. Similarly, the advection speed scales with the SFR surface density as $v\propto \Sigma_{\rm SFR}^{0.4}$ as shown in Fig.~\ref{fig:scaling}(b). However, this relation only holds if the starburst dwarf irregular galaxy IC~10, analysed by \citet{heesen_18c}, is excluded from the fitting. IC~10 has a very high SFR surface density, but only a relatively small advection speed. This outlier may point to the limitations of a scaling with \sfrd. The advection speed scales also with the rotation speed of the galaxy as $v\propto v_{\rm rot}^{1.4}$ (Fig.~\ref{fig:scaling}(c)).  The fact that the advection speed is related to the SFR surface density may be a consequence of a supernovae-driven blast wave \citep{vijayan_20a}. In contrast, for cosmic ray-driven wind models, or for any other wind model, the advection speed is expected to scale with the escape velocity as long as gravity is included  \citep{ipavich_75a,breitschwerdt_91a,everett_08a}, so the scaling with rotation speed is expected as well. Including IC~10 gives an indication that a wind model is preferred, but clearly more dwarf irregular galaxies need to be studied.

\subsection{Accelerated advection speed}
\label{ss:accelerated_advection}
\begin{figure}[tb]
\includegraphics[width=\columnwidth]{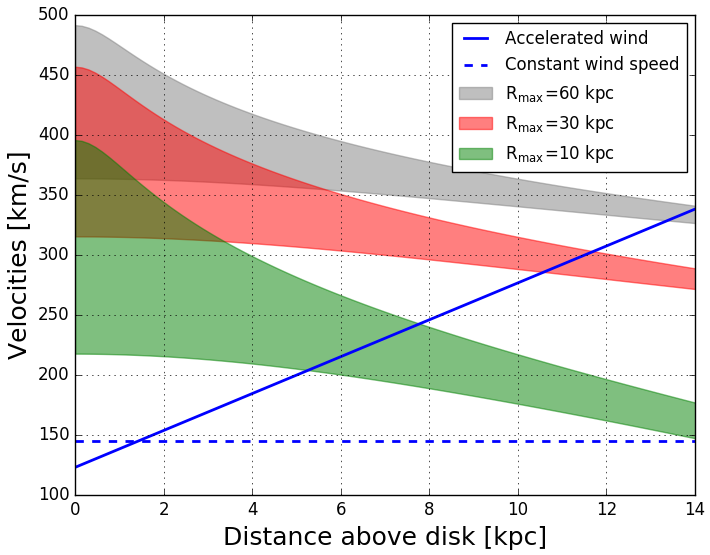}
\caption{Advection speed in NGC~3556 (M~108) as an example for advection-dominated CR$e^{-}$ transport with an accelerating wind. Shaded ares show the expected escape velocity for different dark matter halo distributions. The dashed line shows the best-fitting model with a constant advection speed for comparison. From \citet{miskolczi_19a}} 
\label{fig:m108}
\end{figure}

\begin{figure}[tb]
\includegraphics[width=\columnwidth]{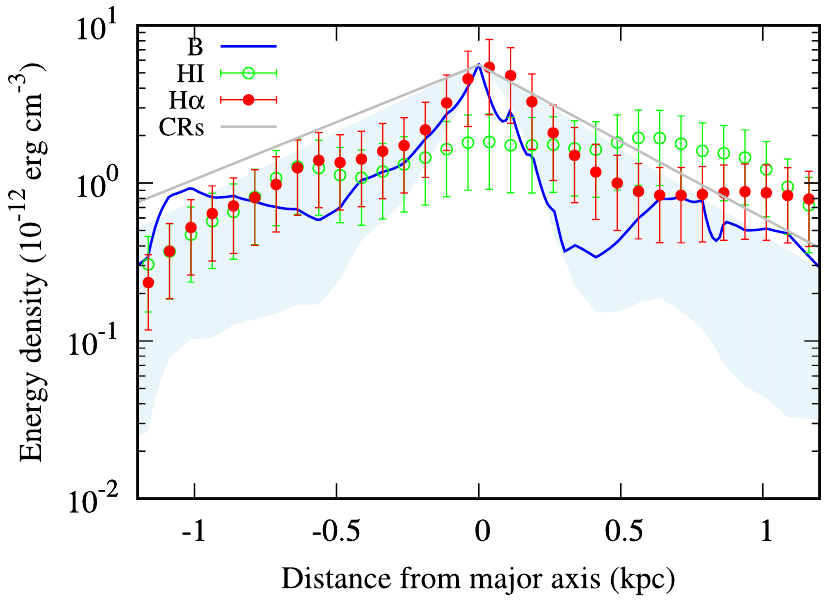}
\caption{Vertical profiles of the energy densities in the dwarf irregular galaxy IC~10 of the magnetic field ($B$), the warm neutral medium (H\,{\sc i}), the warm ionised medium (H\,$\alpha$), and the cosmic rays (CRs). From \citet{heesen_18c}} 
\label{fig:energy}
\end{figure}

With the advent of LOFAR, we are now able to probe the areas in the halo far away from the star-forming mid-plane with height the excess of 10~kpc, where we can probe compatibility of our data with accelerating winds. \citet{miskolczi_19a} have shown that in the galaxy NGC~3556 (M~108) an accelerating wind fits better than advection with a constant wind speed, where they assumed a linearly accelerating wind accelerating from 123~$\rm km\,s^{-1}$ near the mid-plane to 350~$\rm km\,s^{-1}$ at 14~kpc distance (see Fig.~\ref{fig:m108}). This is the first time, where an accelerating wind fits better, whereas with GHz-observations a constant wind speed fits equally well as an accelerating wind \citep{schmidt_19a}. An accelerating wind has the advantage that one can have \emph{energy equipartition} between the cosmic rays and the magnetic field in the halo. For instance, \citet{mora_19a} have shown that a constant advection speed can lead to a divergence between the cosmic-ray energy and the magnetic field of up to a factor of 40 in the halo. For an accelerating wind, cosmic rays can be in equipartition with the magnetic field and possibly even with the warm neutral and the warm ionised gas (see Fig.~\ref{fig:energy}), which is physically more plausible.

%

%

Several possible advection profiles were investigated by \citet{miskolczi_19a}, where they parametrised the advection velocity using equation~\eqref{eq:advection_profile}. For $\beta=1$, the wind is a linearly accelerating, for $\beta=0.5$ the wind acceleration is high near the disc and then tailors off in the halo. They found that $\beta=1$ fits best to the observations. \citet{schmidt_19a} also use a linear advection velocity profile successfully. Hence, a linear advection speed profile appears to be favoured by observations thus far. In \citet{schmidt_19a}, the local advection speed was investigated as well. Surprisingly, the advection is smaller in the centre of the galaxy. This is in contrast to the stronger gravitational acceleration in the centre of the galaxy should lead to higher advection speeds as \citet{breitschwerdt_02a} demonstrated for the case of the Milky Way.

\section{Stellar feedback-driven wind}
\label{s:wind}
Thus far we have used the CR$e^{-}$ as tracers for a galactic wind and neglected the dynamical influence that the cosmic rays have themselves on the wind. Together with the thermal gas they may be able to drive a wind as a result of stellar feedback as is now widely accepted in the literature \citep[e.g.][]{ipavich_75a,breitschwerdt_91a,everett_08a,recchia_16a,mao_18a}. In this section, we present a simple approach that tries to \emph{emulate} such a wind model, but sidestepping the details of cosmic-ray transport which is needed to create such a wind in the first place. For the latter, it is usually assumed that either diffusion or streaming in addition to advection is needed to prevent the adiabatic cooling of the wind. Without such detailed modelling it is not possible to distinguish between the dynamical influence of the thermal and cosmic-ray gas, hence we refer this model to as generic `stellar feedback-driven wind'. Nevertheless, our approach already fulfils some of the requirements we identified in Section~\ref{s:results}:
\begin{itemize}
    \item (i) advection speed is a `wind solution';
    \item (ii) energy equipartition between cosmic rays and the magnetic field;
    \item (iii) linearly increasing advection speed.
\end{itemize}
Assumption (i) is motivated by the fact that a tight correlation between advection speed an escape velocity (i.e. rotation velocity) is observed (Section~\ref{ss:scaling_relations}). Assumption (ii) is made such that energy equipartition is required as suggested by the tight radio--SFR relation (Section~\ref{ss:radio_sfr}). The magnetic fields should be approximately exponential since that is the shape of the vertical intensity profiles (Section~\ref{sss:gaussian_profile_shape}). Assumption (iii) fulfils our finding that linear profiles are well fitting the LOFAR data (Section~\ref{ss:accelerated_advection}). We attempt to meet these requirements with a simple \emph{iso-thermal} wind model.

\subsection{Motivation}
\begin{figure}[tb]
\includegraphics[width=\columnwidth]{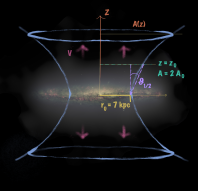}
\caption{Outflow geometry for the stellar feedback-driven wind model. The conical expanding cross-section can be described by the flux tube approximation. From \citet{heald_21a}} 
\label{fig:cone}
\end{figure}

We assume that the cosmic rays are advected in the flow of magnetised plasma, which is directed vertically and expands adiabatically. We use the following functional term for the cross-sectional area:
\begin{equation}
    A(z) = A_0 \left [1 + \left (\frac{z}{z_0}\right )^\beta \right ],
\end{equation}
which describes the `flux tube' geometry. It has been widely used in semi-analytic 1D cosmic ray-driven wind models \citep{breitschwerdt_93a,everett_08a,recchia_16a}. This choice eases the comparison with these aforementioned models. If $\beta=2$ then the model is an expanding cone with a constant opening angle. We may possibly identify these flux tubes with the bubble-like features that definitely play a role as well and disc--halo interface may be more akin to a 'boiling disc' found in radio continuum observations \citep{stein_20a} but also in simulations \citep{krause_21a}. Once these bubbles break out of the thin gaseous disc, the field lines open up and a chimney is formed \citep{norman_89a}. These bubbles and chimneys then may merge and form together a kpc-sized superbubble that expands further into the halo, something that is suggested by the properties of warm dust in the halo \citep{yoon_19a}. The boundary of such a bubble may be related to the X-shaped structures centred on the nucleus, but with footpoints at a galactocentric radius $r_0$. Thus then would define the midplane flow radius, which may be the boundary of this outflow \citep[][see also Fig.~\ref{fig:cone}]{veilleux_21a}. We now also need an equation that governs the magnetic field strength:
\begin{equation}
    B = B_0 \left (\frac{r_0}{r}\right )\times \left ( \frac{v_0}{v}\right ), 
\end{equation}
where $B_0$ is the magnetic field strength in the galactic mid-plane, and $r_0$ and $v_0$ are the mid-plane flow radius and advection speed, respectively. This is the expected behaviour for radial and toroidal magnetic field components in a quasi-1D flow \citep{baum_97a}. Since we do not take rotation into account, we cannot include any dynamical effect that the magnetic field might have on the wind \citep[see][for a simulation of a magnetically driven wind]{steinwandel_20a}. The continuity equation needs to be fulfilled:
\begin{equation}
    \rho v A = \rm const.,
\end{equation}
where $v$ is the advection speed and $\rho$ is the gas density. The momentum conservation is governed by the Euler equation:
\begin{equation}
    \rho v \frac{{\rm d}v}{{\rm d}z} = \frac{{\rm d}P}{{\rm d}z} - g\rho, 
\end{equation}
where $P$ is the combined cosmic-ray and gas pressure and $g$ is the gravitational acceleration. With such a setup, we obtain approximate energy equipartition.  Integrating the Euler equation leads to a wind equation, where we assume for simplicity that the compound sound speed $v_{\rm c}^2=P/\rho$ is constant. It can be shown that the wind velocity profile is in linear approximation:
\begin{equation}
    v = v_{\rm c} \left (1 + \frac{z-z_{\rm c}}{z_{0}} \right ),
    \label{eq:linear_advection_speed}
\end{equation}
where $z=z_{\rm c}$ is the so-called critical point of the wind solution and $v=v_{\rm c}$ is the velocity at the critical point equivalent to the compound sound speed \citep{heald_21a}. This means we can parametrise the wind velocity profile in a linear way as required.

As we do not solve the energy equation explicitly, we have to check whether the energy conservation is indeed fulfilled. This is done via a cloud entrainment factor $\epsilon$, where the total energy flux in the wind is limited by the cosmic-ray luminosity (equation~\ref{eq:cosmic_ray_luminosity}) $L_{\rm CR}=1/2\epsilon \dot M v^2$ with $\dot M$ the global mass-loss rate. This entrainment factor is expected to be of order unity for a cosmic ray-driven wind.
%

\subsection{Application to NGC 5775}
\label{ss:n5775}

 The model is applied to LOFAR 150-MHz and CHANG-ES $1.5$-GHz observations of NGC~5775 \citep{heald_21a}. The data can be indeed well fitted, with a linear acceleration of the advection speed as a result of the wind model (equation~\ref{eq:linear_advection_speed}) and an expanding bi-conical outflow (see Fig.~\ref{fig:cone}). Using the compound sound speed, the thermal electron densities can be calculated. The electron density decreases from a few $10^{-3}~\rm cm^{-2}\,s^{-1}$ by a factor of 10 at the detection limit of the halo at $z\approx 15$~kpc. This phase seems to be most consistent with the hot ionized medium (HIM). There are indications that in certain places the warm ionized medium (WIM) may be entrained in certain places, in particular near the H\,$\alpha$ filaments \citep{tuellmann_00a}. The implied mass-loss rate $\dot M$ is a few solar masses per year. As the advection speeds exceeds the escape velocity at the edge of the halo, it is suggested that the mass is lost entirely from the galaxy. The mass-loss efficiency $\eta=\dot M/SFR$ would then be of order unity. However, we point out that there is substantial uncertainty arising from the outflow geometry and the poorly-understood distribution of ISM material entrained in the vertical flow.

\section{Spectroscopic observations}
\label{s:optical_inferences}

\subsection{Wind speed}
\label{ss:wind_speed}
The arguably most direct way to identify outflows and measure outflow speeds are spectroscopic observations. In the optical wavelength range, the interstellar Na\,{\sc i} absorption line can be used \citep{martin_05a,rupke_05a}, although the drawback of this particular line is that it works only in galaxies at the higher end of the luminosity scale. This limitation was remedied with the Cosmic Origins Spectrograph (COS) aboard the \emph{Hubble Space Telescope} (HST), which made it possible to use ultraviolet absorption lines such as of Si\,{\sc ii} \citep{chisholm_15a} and C\,{\sc ii}, Si\,{\sc iii}, Si\,{\sc iv}, and N\,{\sc ii} \citep{heckman_15a,heckman_16a}. These data trace the warm ionized phase, which is supposed to carry the bulk of the mass in an outflow and so allows us to trace winds in normal star-forming galaxies. This phase can be also seen in emission using the H\,$\alpha$ line, but this again requires high SFRs, so that the galaxies are classified as (U)LIRGs \citep{arribas_14a}.

Our advection speeds increase with the SFR, $\Sigma_{\rm SFR}$ and rotation speed $v_{\rm rot}$, which indicates that they are tracing stellar feedback-driven winds (Section~\ref{ss:scaling_relations}). We now compare the advection speed scaling relations (Table~\ref{tbl:scale}) with the equivalent relation of the gaseous tracers. The UV-absorption line measurements by \citet{chisholm_15a} point to a weak dependence of the outflow speed with the SFR of $v\propto SFR^{0.08-0.22}$ and similarly with the rotation speed of  $v\propto v_{\rm rot}^{0.44-0.87}$. In contrast, \citet{heckman_16a}, also using UV-absorption lines, find much stronger dependencies with $v\propto SFR^{0.32\pm 0.02}$ and $v\propto v_{\rm rot}^{1.16\pm 0.37}$ \citep[see also][]{heckman_15a}. \citet{martin_05a} used Na\,{\sc i} and K\,{\sc i} absorption lines in ultra-luminous infrared galaxies and found $v\propto SFR^{0.35}$. 

On the subject of whether the wind speed depends on \sfrd, the literature is even more divided. \citet{chisholm_15a} did find no notable correlation, whereas \citet{davies_19a} claim a strong correlation of $v\propto \Sigma_{\rm SFR}^{0.34\pm 0.10}$. Notably, the sample of \citet{davies_19a} contains mostly star bursts with \sfrd $= 0.1$--1~\usfr, whereas the sample by \citet{chisholm_15a} covers also lower values of \sfrd. \citet{heckman_16a} claimed a correlation of $v\propto \Sigma_{\rm SFR}^{0.34}$ up to a value of 100~\usfr, flattening out at even higher values. In Fig.~\ref{fig:scaling}, we compare the UV measurements of \citet{heckman_15a} with our advection speeds. In general, we find a good agreement with their wind speeds as function both of the SFR and rotation speed, although the scatter is fairly large for the UV measurements. For the comparison with the SFR surface density, there is no such good agreement, with our winds happening at much lower values of \sfrd. In part this may be explained by our different definition of \sfrd, which employs the full extent of the star-forming disc whereas \citet{heckman_15a} use an effective (half-light) star-forming disc radius.

\subsection{Mass loading}
\label{ss:mass_loading}
The mass-loading factor is defined as $\eta=\dot M/SFR$, where $\dot M$ is the mass-loss rate. The mass-loading factor is predicted to increase strongly with decreasing rotation speed, so that in dwarf galaxies the mass-loading factor could easily exceed unity, whereas in Milky Way-type $L_\star$ galaxies, the factor is of order unity. \citet{chisholm_17a} parametrised the mass-loading factor as:
\begin{equation}
    \eta = 1.12 \pm 0.27 \left (\frac{v_{\rm rot}}{100~\rm km\,s^{-1}} \right )^{-1.56\pm 0.25},
    \label{eq:mass_loss_rate}
\end{equation}
using UV-absorption line studied of outflows. Similarly, \citet{heckman_16a}, also using UV-absorption line studies, found a slightly flatter dependency of $\eta\propto v_{\rm rot}^{-0.98}$. While we have not applied our stellar feedback-driven wind model (Section~\ref{s:wind}) to a sample yet, we can use the theoretical expectation of a similar cosmic ray-driven wind model of $\eta\propto v_{\rm rot}^{-5/3}$ \citep{mao_18a}, which gives quite reasonable agreement. It is also encouraging that our one data point for NGC~5775 (Section~\ref{ss:n5775}) predicts a mass-loading factor of order unity, which is in good agreement with equation~\eqref{eq:mass_loss_rate}.

\subsection{Wind velocity profile}
\label{ss:wind_velocity_profile}
Wind velocity profile measurements are only few and far between since it requires spatially resolved line observations. The wind velocity profiles from the optical measurements look significantly differently than linear acceleration, where the acceleration happens close to the disc and converges quickly \cite{chisholm_16a}. Notably, the acceleration happens already largely within 1~kpc from the star burst region. \citet{chisholm_16a} attribute this velocity profile to either radiation pressure or cosmic-ray pressure, assuming that the accelerating force falls of with distance squared. There are a handful of other galaxies where the wind velocity profile has been measured such as in NGC~253 \citep{westmoquette_11a}, where outflow speeds of a few hundred \uvel\ are found within a few 100~pc from the disc and which increase  linearly with height.

Our radio haloes may require acceleration in particular if the lateral expansion needs to be limited as the morphology of the radio haloes suggests. On the other hand, the wind models such as of \citet{chevalier_85a} even with the inclusion of cosmic rays \citep{samui_10a,yu_20a} all predict rapid acceleration near the disc even when adopted to the flux tube geometry \citep{heald_21a}. Hence, the jury is still out whether the wind velocity profiles are more in agreement with a linear acceleration across the size of the halo ($\sim$10~kpc), possibly extending even further, as some wind models predict that do not include an extended area of mass-loading but inject all energy at $z=0~\rm kpc$ \cite{breitschwerdt_91a,everett_08a,recchia_16a}.  While using the radio spectral index is a rather indirect way of measuring the velocity profile and subject to assumptions about the magnetic field, some form of acceleration seems to be most plausible as it is also the result of any stellar feedback-driven wind model (Section~\ref{s:wind}).

\subsection{Outflow size}
\label{ss:outflow_size}

The outflow size in most absorption line studies is only a few kpc at most \citep{heckman_16a}, whereas radio haloes are indicative of galaxy-wide outflows with radii typically a few kpc. Although the boundary of radio haloes is poorly defined, but the size of the haloes is typically comparable to the size of the star-forming disc \citep{dahlem_06a}. The connection of the galaxy-wide outflows with nuclear star bursts is rather uncertain \citep{westmoquette_11a}. A case in point is NGC~253, which does have a nuclear star burst with \sfrd $\sim$ 1~\usfrd, which shows a well-defined nuclear outflow \citep{heesen_11a}. The same galaxy has also a galaxy-wide advective radio halo \citep{heesen_09a} and an X-ray halo indicating a galaxy-wide outflow as well \citep{bauer_08a}. It is possible that the `down the barrel' optical and UV spectroscopic surveys do overlook the larger size of the outflow region due to sensitivity issues since a broad component emission or absorption line has to be identified. 

When other measurement are used such integral field unit (IFU) spectroscopy, the size of the haloes are much larger in width. In the SAMI data of \citet{ho_16a}, the velocity field is widely asymmetric in galaxies indicating a larger outflow size.  In the CALIFA sample, \citet{lopez_coba_19a} identified outflows with increasing line ratios such as [N\,{\sc ii}]/H$\alpha$ along the semi-major axis. Such increasing ratios are consistent with shock ionization in galactic outflows. Again, the morphology points to galactic outflows.

\subsection{Outflow threshold}
\label{ss:outflow_threshold}

The existence of a minimum value for the star-formation rate surface density was first posed by \citet{rossa_03b}, who studied the extra-planar diffuse ionized gas (eDIG) in edge-on galaxies. Their value is \sfrd $= 2\times 10^{-3}$~\usfrd, which was then corroborated by \citet{tuellmann_06a} who studied extra-planar hot ionized gas via X-ray emission. In most galaxies, there is no extended eDIG emission detected below this threshold, and if there is a detection, the dust temperature is significantly higher. This threshold is much lower than the canonical threshold for galactic winds by \citet{heckman_00a}, who suggested \sfrd $\approx 10^{-1}$~\usfrd. Galaxies exceeding this value are commonly referred to as `superwind' galaxies and are known to have extensive X-ray haloes \citep{strickland_04a}. More recent observations have shown this outflow threshold to be potentially much lower, as for instance the detection of a superbubble of warm dust in NGC~891 with a local \sfrd\ of $0.03~$\usfrd\ suggests \citep{yoon_19a}. It is probably the local value of \sfrd\, which needs to be $\sim 10^{-2}$~\usfrd\ to allow the formation of chimneys facilitating outflows. The chimneys would form at spiral arms and predominantly at smaller galactocentric radii allowing an outflow in the inner parts of the galaxy \citep[see instructive simulations by][]{krause_21a}.

\citet{ho_16a} suggests much lower values of globally averaged star formation rates with \sfrd $=SFR/(\pi r_{\rm e}^2)$ with values of $10^{-3}$--$10^{-1.5}$~\usfrd, where $r_{\rm e}$ is the effective radius. \citet{lopez_coba_19a} suggest \sfrd $> 10^{-2}$ \usfrd\ in conjunction with a centrally concentrated gas distribution. Clearly, this value depends on the detection method. The radio continuum method suggests also a low threshold of around the value as for the existence of eDIG. This threshold is identified by the transition from diffusion-dominated haloes to advection-dominated ones (Section~\ref{ss:profile_shape} and Fig.~\ref{fig:scaling}(b)). That this property of radio haloes fits to the optical observations raises the possibility that the vertical profile of radio haloes (Gaussian and exponential, Section~\ref{ss:profile_shape}) and the multi-component fitting (Section~\ref{sss:multi_component_radio_disc}) allow us already to distinguish between galaxies with outflows and those that do not have outflows.

\subsection{Cosmic-ray calorimetry}
The detection of $\gamma$-rays from star-forming galaxies with \emph{Fermi} has allowed us to compare the $\gamma$-ray luminosity with expectations from cosmic-ray calorimetry \citep{ackermann_12a}. \citet{lacki_11a} show that the star-burst nuclei of NGC~253 and M~82 are closest to calorimetry, although there is some uncertainty with regards to the GeV-emission from secondary CR$e^{-}$. \citet{yoast_hull_13a} showed that M~82 is a good electron calorimeter but not proton calorimeter, while for NGC~253 the situation is more complex \citep{yoast_hull_14a}. It has been recently suggested by \citet{hopkins_20a} that the other galaxies at lower SFRs are only poor cosmic-ray proton calorimeters. This requires a fast escape of cosmic rays either by diffusion, as facilitated with a high diffusion coefficient \citep{hopkins_20a}, or via a galactic wind.

\subsection{Extra-planar gas}
The extra-planar gas comprises several phases, the warm neutral medium traced by H\,{\sc i}, the WIM traced by H\,$\alpha$, and the HIM traced by X-ray emission. The HIM as traced by X-ray emission has electron density scale heights between 4 and 8~kpc with exponential profiles preferred over Gaussian or power-law profiles \citep{strickland_04a,hodges-kluck_13a}. These data would be in approximate agreement with what is expected for an outflow of a hot wind. In contrast, the WIM as traced by H$\alpha$ emission has much smaller scale heights of $\sim$1~kpc \citep{dettmar_06a}, although in places there can be filaments with much larger scale heights of 3--5~kpc \citep{boettcher_13a}. These profiles can be again approximated by exponential functions. The atomic gas as traced by H\,{\sc i} emission has a large variety of scale heights \citep{zschaechner_15a}, sometimes with both a thin and thick in the order of between a few 100~pc and a few kpc. 

Because both the H\,{\sc i} and H\,$\alpha$ are line emissions, one can measure the rotation speed of edge-on galaxies as function of height. It is observed that the rotation speed of the gas decreases approximately linearly with height. This is referred to as `rotational lag' and has an amplitude of 5--20~$\rm km\,s^{-1}\,kpc^{-1}$ \citep{heald_06a,zschaechner_15a}. 

\section{Inferences from theory}
\label{s:inferences_from_theory}

\subsection{Cosmic ray-driven wind models}

For cosmic rays to able to drive a wind, they have to be effectively confined in the galaxy for some time. We recall that the cosmic-ray mean free path is only  a few pc, so that galaxies are effectively optically thick for cosmic rays. The cosmic rays then transfer a small part of their momentum and energy on the gas every time they interact via Alfv\'en waves, which they generate themselves via the streaming instability \citep[`self-confinement' picture][]{zweibel_13a}. The 1D cosmic ray-driven wind models by \citet{breitschwerdt_91a} and \citet{everett_08a} showed that cosmic rays streaming along the magnetic field lines can compensate for the adiabatic cooling of the wind fluid and so the compound sound speed increases slightly in the halo. This is required for a wind solution to go through the critical point where the gravitational acceleration is approximately constant as in the case for a galaxy halo \citep{mao_18a}. The wind velocity profiles are approximately linear before the speed converges to a few times the rotation speed \citep{everett_08a}. One of the limitations of these wind models is that all the energy and mass are injected at $z=0$, which is not very physical. The widely used analytical wind model of \citet{chevalier_85a} has a driving region where the energy and momentum are injected, which is more realistic. This model was extended by \citet{samui_10a} to include cosmic rays, which shows rapid acceleration in the driving region and nearly constant velocity at larger radii. As the driving region is rather confined to the disc plane with a height of $\sim 100$~pc, similar to the gaseous scale height of the warm neutral medium, the acceleration will be small at heights $\gtrsim$1~kpc.

In a steady state and without considering internal losses, the cosmic rays would transfer a fraction of their luminosity on the total energy flux in the wind, so $L_{\rm CR}\approx 1/2(\epsilon/0.5)\dot M v^2$, where $\epsilon$ is an efficiency factor. As we shown in Section~\ref{s:wind}, this is consistent with a stellar feedback-driven wind model in NGC~5775. A consequence of cosmic ray-driven winds is that the hydrodynamical equilibrium state of galaxies will be affected by the cosmic rays \citep{crocker_20b}. Another property of cosmic ray-driven wind models, which include gravity, is that the wind velocity scales linearly with the rotation speed in the galaxy \citep{ipavich_75a}. Such a behaviour is also expected for `momentum-driven' winds for which radiative cooling is important \citep{murray_05a,veilleux_20a}. The advection speed scaling relation (Section~\ref{ss:scaling_relations}) with the rotation speed is in good agreement with such an expectation. 

%
\begin{table*}[tb]
\small
\caption{Cosmic ray (CR)-driven 1D wind models} 
 \label{tbl:cr1d}
\begin{tabular}{lc ccccc}
 \tableline  
  Model & Driving & Diffusion & Streaming & Gravity & Rotation & Reference \\
  \tableline  
  \tablenotemark{a}Constant wind speed & N/A & $\times$ & $\times$ & $\times$ & $\times$ & \citet{heesen_18b} \\
  \tablenotemark{a}Accelerated wind & N/A & $\times$ & $\times$ & $\times$ & $\times$ & \citet{miskolczi_19a} \\
  \tablenotemark{b}Iso-thermal wind    & hybrid & $\times$ & $\times$ & \checkmark & $\times$ & \citet{heald_21a} \\
  CR-driven wind with streaming & hybrid & $\times$ & \checkmark & \checkmark & $\times$ & \citet{everett_08a} \\
  CR-driven wind with diffusion  & hybrid & \checkmark & \checkmark & \checkmark & $\times$ & \citet{recchia_16a} \\
  CR-driven wind with rotation  & hybrid & $\times$ & \checkmark & \checkmark & \checkmark & \citet{zirakashvili_96a} \\
  CR-driven wind w.\ rot. \& diff.\  & hybrid & \checkmark & \checkmark & \checkmark & \checkmark & \citet{ptuskin_97a} \\
  \tableline 
 \end{tabular}
%
 \tablenotetext{a}{CR$e^{-}$ are only used as tracers with no dynamical innfluence of the cosmic rays}
 \tablenotetext{b}{CR$e^{-}$ are used as tracers with a simplified iso-thermal wind model that includes cosmic-ray pressure (Section~\ref{s:wind})}
 \tablecomments{Hybrid driving means both a dynamical influence of the thermal gas (HIM) and the cosmic rays}
 \end{table*}

\subsection{Simulations}

Simulations have shown how important cosmic rays are in order to create galaxy-wide outflows. There are now many MHD simulations available that include cosmic rays such as the one of \citet{girichidis_18a}. They showed that cosmic ray-driven outflows are significantly cooler with a large fraction of the gas staying at $10^4$~K rather than $10^6$~K for the thermally driven case. The reason that cosmic rays are so effective at driving galactic winds is that they can diffuse out of star-forming regions and then create a `background sea' with a pressure gradient on kpc-scales \citep{salem_14a}. This gradient then can lift the gas into the halo. The transport of cosmic rays is important as if the cosmic rays are just advected with the gas, they act only as an additional pressure component and so the gaseous disc is `puffed up', but no outflow is created \citep{ruszkowski_17a}.

The cosmological simulations by \citet{pakmor_16a} showed the influence of anisotropic cosmic-ray diffusion. Galaxies with spiral magnetic fields created in part by differential rotation, can store cosmic rays for longer. While a wind develops both in the case of anisotropic and isotropic diffusion, the isotropic diffusion suppresses the magnetic field amplification in the disc. Again, without diffusion, a wind does not form at all. In \citet{jacob_18a}, simulations of dwarf galaxies show spherical winds with slow speeds of 20~\uvel, whereas galaxies with higher masses have more bi-conical outflows with higher speeds of 200~\uvel. Interesting, for galaxies in excess of a virial mass of $10^{11.5}~\rm M_{\odot}$ do not form cosmic ray-driven winds beyond the virial radius. For all cases, the outflow speed is in good agreement with the escape velocity near the galactic mid-plane. However, their wind speeds seem to be always on the low side when compared with optical observations and our data.

\subsubsection{Wind velocity profiles}

The simulations by \citet{girichidis_18a} showed approximately linearly accelerating mass-weighted velocity profiles. They simulated only the first 2~kpc near the galactic mid-plane, so that it is not clear whether the escape velocity is reached. The authors point out that further acceleration is expected in the halo. The outflow speeds they find are $\sim$50~\uvel, so significantly lower than what we measure near the disc. In MHD simulations of isolated galaxies with cosmic rays, \citet{jacob_18a} found that the vertical velocity profiles are in good agreement with the wind model of \citet{chevalier_85a}, with rapid acceleration near the disc and then a nearly constant velocity in the halo.

\subsubsection{Cloud entrainment}

The entrainment of clouds is important to load the wind, which contains mostly of the HIM, with further mass. \citet{banda_20a} showed that the hot wind is able to accelerate clouds at a speed comparable to the escape velocity within the first 1~kpc away from the the disc. These clouds form a `mist' of WIM that can be further accelerated in the wind. The wind is formed this close to the disc as expanding superbubbles which are particularly good in converting the kinetic energy of SNe into thermal energy, i.e. the thermalization efficiency is quite high with a few 10 per cent \citep{sharma_14a}. What appears to be important is that these clouds can then be considered as the starting point for the reference level at 1~kpc where we start to see advection-dominated haloes. It appears hence as possible that these clouds are advected as a thin mist with the fast flow of the HIM and traced by the CR$e^{-}$. 

As noted in Section~\ref{ss:wind_speed}, there is a good agreement between the scaling relations as found for the advection speeds and the cool neutral and warm ionized outflows as measured from UV interstellar absorption lines. This is the case for both the magnitude of the outflow velocity as well as the scaling relations slopes. This agreement, while assuming that radio traces rather the HIM (Section~\ref{ss:n5775}), needs entrainment of clouds in order to get similar wind speeds for the clouds.

\section{Missing physics}
\label{s:missing_physics}

We have now approached a phase where we have demonstrated that radio continuum observations can give us a complementary view on galactic winds. What we have not yet been able to do though is to identify the mechanisms that allow us to probe the workings of a cosmic ray-driven galactic wind. In our iso-thermal wind model we have only \emph{assumed} that the cosmic rays do not cool adiabatically and the sound speed is constant (Section~\ref{s:wind}). This can be achieved both by either cosmic-ray streaming along magnetic field lines or anisotropic diffusion. This has been achieved already with 1D models in the literature, so their application remains to be carried out in future work. In Table~\ref{tbl:cr1d}, we present an overview of our pure phenomenological models and the 1D wind models from the literature. We now discuss the main physical effects.

\subsection{Cosmic-ray streaming}

The vertical spectral index profiles are well fitted with our advection models, although our concave model profiles may still be improved in order to fit the data better \citep{miskolczi_19a}. This would require a faster CR$e^{-}$ bulk speed without the change in magnetic field as demanded by the continuity equation and energy equipartition. Such a change may hence be in agreement with cosmic-ray streaming which does change the CR$e^{-}$ bulk speed. As it was also suggested, the streaming speed can be even a few times the Alfv\'en speed, when the cosmic rays decouple from the magnetic field such as when neutral atoms suppress Alfv\'en waves \citep{ruszkowski_17a}. This raises the possibility to detect a particularly fast cosmic-ray bulk speed in areas with an excess of H\,{\sc i} emission.

An alternative suggestion is to measure the velocity differential between the CR$e^{-}$ bulk speed and the ionized gas in the outflow. For edge-on galaxies, the line-of-sight velocity is of course fairly small depending on the outflow opening angle. The entrained clouds should soon reach the speed of the HIM, so that we can use these clouds as a tracer for the outflow speed \citep{banda_20a}. The global wind speeds appear to be in good agreement with what has been measured for the WIM using optical and UV spectroscopy (Section~\ref{ss:wind_speed}), but there may still be local effects in particular where the magnetic field has a strong vertical component \citep{tuellmann_00a}.

\subsection{Cosmic-ray diffusion}

Near the galactic mid-plane, cosmic-ray diffusion is important. The high-angular resolution images resolved on a 1-kpc scale resolve the thin radio disc. Hence, the escape of the cosmic rays is governed by the superposition and advection. Since we do not resolve this region well with our observations, we have neglected its influence. However, as \citet{breitschwerdt_93a} and \citet{recchia_16a} have shown, this region is quite important for the launching of the wind and also for the cosmic-ray spectrum \citep{ptuskin_97a}. Obviously, the escape of CR$e^{-}$ would have a bearing also on the integrated radio spectral index, which we can measure now over several decades in frequency. As pointed out before, while we may neglect diffusion in comparison to advection for simply measuring the transport of CR$e^{-}$ (Section~\ref{sss:cre_transport_length}), from a theoretical point of view we cannot trace the makings of a cosmic ray-driven wind without taking diffusion into account.

\subsection{Rotation}

We have neglected the influence of the magnetic pressure on the outflow velocity. This is the case if the outflow follows the magnetic field lines, so that lines of magnetic force are parallel to the wind velocity. Obviously, this will have some consequence for the outflow geometry which we can potentially test with linear polarisation measurements. Due to the superposition of the azimuthal velocity and vertical velocity, the magnetic field lines would wind up in a helical shape in the halo. This may be potentially observable as a rotation measure signal. So far linear polarisation measurements have shown X-shaped magnetic fields but only little or no large-scale rotation measure signal in the halo \citep{soida_11a}.

What then also needs to be taken into account is a proper treatment of the angular momentum in the outflow. This will change the wind solutions as \citet{zirakashvili_96a} have demonstrated. They also found that the dynamics of the ionized gas in the halo is changed since some of the angular momentum is transferred to this gas.

\section{Summary}
\label{s:summary}

As we have demonstrated in this review, radio continuum observations open up a new window on cosmic-ray transport in nearby galaxies. These observations allow us to calibrate the influence that cosmic rays have on galactic winds, a process that shapes and influences galaxy evolution in a unique way. Cosmic rays are transported by diffusion, advection, and streaming, which all contribute to a different degree. Since galaxies have complex magnetic field configurations, they are effectively optically thick to the scattering of cosmic rays. With radio continuum observations we trace cosmic-ray electrons in GeV-energy range, corresponding to the peak of the cosmic ray energy density in the spectrum of protons and heavier nuclei.  In summary, our main results are as follows:
\begin{enumerate}
    \item Diffusion coefficients for GeV-cosmic rays are of order $10^{28}$~\udif, with either little or no energy dependence (Section~\ref{ss:diffusion_coefficients});
    \item Gaussian radio haloes are diffusion-dominated and are found in galaxies with no winds, whereas exponential radio haloes are advection-dominated indicative of winds (Section~\ref{ss:profile_shape});
    \item Advective radio haloes are predominant in galaxies with \sfrd$\geq 2\times 10^{-3}$~\usfrd\ (Section~\ref{sss:gaussian_profile_shape}); this suggests that there is a \sfrd-threshold value for galactic winds (Section~\ref{ss:outflow_threshold});
    \item The advection speed scales with SFR, \sfrd, and $v_{\rm rot}$ (Section~\ref{ss:scaling_relations}), corroborating stellar feedback as the cause for radio haloes;
    \item The high advection speeds, comparable to the escape velocities, suggest that the gas can escape into the CGM and contribute to the escape of baryons and metals;
    \item The advection speed scaling relations are in good agreement with what has been measured using optical and UV spectroscopy, further corroborating galactic winds and radio haloes have a common cause (Section~\ref{ss:wind_speed}); 
    \item A stellar feedback-driven wind model suggests that the hot ionized medium is the main wind fluid (Section~\ref{s:wind}). If cosmic rays are driving the wind, the mass-loading factor could be of order unity for Milky Way-type galaxies. However, due to the uncertain geometry and entrainment of warm ionized and cold neutral medium clouds, this mass-loss rate might be substantially different. 
\end{enumerate}

There are a few caveats relevant to these conclusions, however. One of them is that it can be difficult to distinguish advection and diffusion based purely on the radio spectral index \citep{stein_19b}. It is thus possible that galaxies near the diffusion--advection boundary may be mis-classified. Another uncertainty is the unknown contribution from cosmic-ray streaming in edge-on galaxies, which may lead to advection speeds overestimating wind velocities. Hence, in the future a better modelling of streaming would be required, incorporating it into the stellar feedback-driven wind model (Section~\ref{ss:cosmic_ray_streaming}). In face-on galaxies, a better modelling of cosmic-ray diffusion and streaming while accounting for the escape of CR$e^{-}$ would be necessary in order to affirm measurements of the diffusion coefficient and to be able to distinguish between these transport modes (Section~\ref{s:face_on_galaxies}).

Our long-term goal is the aim to inform cosmological simulations of galaxies, which have to build in these kind of physics as part of `subgrid' models \citep{vogelsberger_20a}. With new radio facilities now producing many observational data sets we can test these models, we expect that important input will come from observations for the foreseeable future.


%
%

%
%

%

%
%

\section*{Data availability}
All data generated or analysed during this study are included in this published article.

\section*{Statements and Declarations}
The authors have no competing interests to declare that are relevant to the content of this article.

%
\acknowledgments
We would like to thank the editors of this Topical Collection, Manami Sasaki, Ralf-J\"urgen Dettmar and Julia Tjus, for the invitation to write this review article. We also would like to thank the anonymous referee for an insightful report that helped to improve this paper. We thank our colleagues for providing important contributions to this article. In particular, we would like to thank Arpad Miskolczi, Yelena Stein, Philip Schmidt, Sarrvesh Sridahr, and George Heald. They all helped a lot with studying the radio continuum observations and applying the {\sc spinnaker} models to them. Arpad Miskolczi is also thanked especially for developing {\sc spinteractive}, which made the application of these models so much easier.


%
\bibliographystyle{spr-mp-nameyear-cnd}  
\bibliography{review}                

%

\end{document}